\documentclass[%
 reprint,
nofootinbib,
 amsmath,amssymb,
 aps,
showkeys
]{revtex4-2}

\usepackage{graphicx}
\usepackage{subcaption}
\usepackage{dcolumn}
\usepackage{bm}
\usepackage{xfrac}
\usepackage{hyperref}
\hypersetup{colorlinks=true,linkcolor=red,citecolor=blue,filecolor=blue,urlcolor=blue}
\usepackage{booktabs}
\usepackage{orcidlink}

\usepackage{multirow}

\usepackage{aasmacros}

\usepackage{url}

\begin{document}
\defcitealias{finke_modeling_2022}{Finke22}
\defcitealias{haardt_radiative_2012}{CUBA}

\graphicspath{{./}{Images/}}

\title{Novel bounds on decaying axionlike particle dark matter from the cosmic background}

\author{Sara Porras-Bedmar$^1$\,\orcidlink{0009-0001-3512-2628}}
\email{sara.porras.bedmar@uni-hamburg.de}
\author{Manuel Meyer$^2$\,\orcidlink{0000-0002-0738-7581}}
\author{Dieter Horns$^1$\,\orcidlink{0000-0003-1945-0119}}
\affiliation{
 $^1$Universität Hamburg, Luruper Chaussee 149, D-22761 Hamburg, Germany\\
 $^2$CP3-Origins, University of Southern Denmark, Campusvej 55, DK-5230 Odense M, Denmark}



\date{\today}

\begin{abstract}
The cosmic background (CB) is defined as the isotropic diffuse radiation field with extragalactic origin found across the electromagnetic spectrum. 
Different astrophysical sources dominate the CB emission at different energies, such as stars in the optical or active galactic nuclei in x rays. Assuming that dark matter consists of axions or axionlike particles with masses on the order of electron volts or higher, we expect an additional contribution to the CB due to their decay into two photons. Here, we model the CB between the optical and x ray regimes, and include the contribution of decaying axions. Through a comparison with the most recent direct and indirect CB measurements, we constrain the axion parameter space between masses $0.5 \mathrm{eV} - 10^7\mathrm{eV}$ and improve previous limits on axion-photon coupling derived from the CB by roughly an order of magnitude, also reaching the QCD band. We further study the contribution of axions decaying in the Milky Way halo and characterize the axion parameters that would explain the tentative excess CB emission observed with the Long Range Reconnaissance Imager instrument on-board the New Horizons probe.

\end{abstract}

\maketitle

\section{Introduction\label{section:Intro}}

The $\Lambda$CDM cosmology relies on the existence of nonbaryonic cold dark matter (DM), which  makes up around 85\,\% of the matter content in the Universe~\cite{2020plank}.
So far, evidence for dark matter comes solely from its gravitational interaction revealing itself in astrophysical observations on different scales~\cite{ReviewParticlePhysics}. One possibility is that dark matter is made up from so-far undetected fundamental particles. Axions are one such candidate~\cite{preskill_cosmology_1983, dine_not-so-harmless_1983}.

The axion is a proposed pseudo-Nambu–Goldstone boson that arises from the Peccei–Quinn solution to the strong \textit{CP} problem in QCD, thus explaining the nonobservation of the electric dipole moment of the neutron~\cite{peccei_cp_1977,peccei_constraints_1977}. Axions are predicted to couple to photons, with their coupling being proportional to the axion mass~\cite{2018PrPNP.102...89I}. Axionlike particles (ALPs) are an extension of axions that do not solve the strong \textit{CP} problem and whose mass and coupling are independent parameters.
Just like axions, ALPs are a well-motivated candidate to explain DM~\cite{2012Arias}.

The search for axions and ALPs has significantly intensified over recent years: direct searches are carried out with haloscopes~\cite{salemi_search_2021, the_admx_collaboration_squid-based_2010,adair_search_2022,alesini_search_2022} and helioscopes~\cite{cast_collaboration_new_2017,betz_first_2013,ballou_new_2015,della_valle_pvlas_2016,kirita_search_2022}, axion and ALP production is tested in laboratories~\cite{atlas_collaboration_measurement_2021,belle_ii_collaboration_search_2020,cms_collaboration_evidence_2019,aloni_photoproduction_2019,astier_search_2000, knapen_searching_2017}, and in astrophysical environments~\cite{kohri_axion-like_2017, caputo_low-energy_2022,buen-abad_constraints_2022,li_limits_2021,dessert_no_2022,calore_3d_2022}, sometimes assuming ALPs are DM~\cite{cadamuro_cosmological_2012,janish_hunting_2023,wadekar_strong_2022,grin_telescope_2007,todarello_robust_2023,carenza_probing_2023,calore_constraints_2023}.
Many of these searches make use of the coupling of ALPs to photons, $\mathcal{L}_{a \gamma} = -\frac{1}{4} g_{a \gamma} F_{\mu \nu} \widetilde{F}_{\mu \nu} a$, with photon coupling $g_{a \gamma}$, the ALP field strength \textit{a}, and $F_{\mu \nu}$ the electromagnetic field tensor (and its dual $\widetilde{F}_{\mu \nu}$). This coupling can lead to photon-ALP conversion in magnetic fields or ALP decay into two photons. We focus on an indirect astrophysical method assuming ALPs as DM that could contribute to the cosmic background (CB).

The CB is defined as the homogeneous and isotropic diffuse radiation field with extragalactic origin. It covers the entire electromagnetic spectrum, from radio waves to $\gamma$ rays, so different detection techniques and observations are necessary to trace it. Recent reviews and data compilations of CB measurements are available from Refs.~\cite{hill_spectrum_2018, Biteau_The_MM_EGAL_spectrum}. Various astrophysical processes contribute to the CB at different wavelengths. The potential decay of ALPs into two photons would add to the CB intensity, its exact spectral shape given by the ALP mass and photon coupling strength, and its energy shifted by cosmological expansion. 
Limits over a broad range of mass and coupling constants were derived in Ref.~\cite{cadamuro_cosmological_2012}.

The cosmic optical background (COB) has a special importance regarding ALP searches. Given the relation between axion mass and decay wavelength, the decay from the QCD axion would contribute in this regime.
The New Horizons probe (NH) has been traveling though the outskirts of the Solar System, and has taken various measurements of the optical CB through the years, some of them over 51$\,$A.U. away from Earth~\cite{lauer_anomalous_2022, 2023lorri}. These measurements are relevant because the zodiacal light~\cite{1998HauserZL}, solar light scattered by dust, is less prevalent in the outer region of the Solar System.
Zodiacal light is a bright foreground in the optical and infrared bands, and its modeling currently is not sufficiently accurate to subtract it reliably from the measurements to obtain the comparably weak COB.
This makes the NH measurements unique, since it is the only occasion where we have subdominant interplanetary dust contamination in an optical observation.

After the extraction of the residual foreground emission,
the authors of Ref.~\cite{2023lorri} found a $>5\sigma$ mismatch in intensity between the NH measurement and the galaxy counts calculated by previous works. 
Galaxy counts are defined as the minimum amount of intensity that galaxies would contribute to the COB~\cite{driver_measurements_2016,windhorst_jwst_2023}, and they can be understood as COB lower limits.
This excess has been interpreted in terms of axion decay~\cite{bernal_cosmic_2022,bernal_seeking_2022}. 
However, no other COB measurements were included in their study. Any axion or ALP interpretation must also be consistent with constraints on the COB at other optical wavelengths.
Furthermore, a recent analysis of a larger dataset obtained with Long Range Reconnaissance Imager (LORRI) suggests a much reduced excess compatible at the $1.4\,\sigma$ level with Galaxy number counts~\cite{2024postmann}.

Other spectral regimes are also interesting to study. In this work we focus on the broad wavelength range between $10^{-6}$ and $10\, \mu \mathrm{m}$ or from x rays to optical wavelengths.
The three regimes included here are the cosmic x-ray, ultraviolet and optical backgrounds (CXB, CUB and COB, respectively).
Longer wavelengths would mean entering the infrared regime or CIB. The infrared part of the CB requires additional assumptions about the dust population~\cite{saldana-lopez_observational_2021,finke_modeling_2022}, which we leave for future study.
Smaller wavelengths mean studying the $\gamma$-ray background (CGB). This would entail modeling the different populations of active Galactic nuclei (AGN) and accounting for the absorption of $\gamma$ rays in the intergalactic medium. We also leave this for future work.

In summary, the goals of our study can be summarized as follows:
\begin{enumerate}
    \item Investigate the interpretation of the NH measurement through axion decay in light of the excess reported in Ref.~\cite{2023lorri} and the updated result in Ref~\cite{2024postmann}.
    For this we develop our own model of the COB. 
    \item Derive novel constraints on DM axions and ALPs through their contribution to the CB from optical to x rays taking astrophysical emission into account.
    \item Investigate the decay contribution from the local Milky Way DM halo. A high DM local density entails an extra contribution to CB measurements.
\end{enumerate}

Throughout this work, we use $\Lambda$CDM cosmology with $\Omega_{\Lambda}=0.7, \Omega_\mathrm{m}=0.3$ and $\Omega_\mathrm{baryons}=0.045$ (as the fraction of matter made up from baryons)~\cite{2020plank}. 
We take a value for the Hubble constant of $H_\mathrm{0} = 70\,\mathrm{km}\,\mathrm{s}^{-1}\,\mathrm{Mpc}^{-1}$, in order to be consistent with previous analyses whose results are used here.

The paper is structured as follows: 
in Sec.~\ref{section:measurements} we describe the observational data that we are going to work with.
Then, in Sec.~\ref{section:theoreticalBackground} we describe the theoretical background of astrophysical processes that form the CB and its modeling. In addition to the CB, we mainly include the axion/ALP decay, the DM host halo, and stellar populations.
In Sec.~\ref{section:results} we present our results from modeling and the constraints to axions.
The results are discussed in Sec.~\ref{section:discusssion} and we conclude in Sec.~\ref{section:conclusions}.

\section{Measurements of the CB} \label{section:measurements}

\begin{figure*}[ht!]
    \centering
    \includegraphics[width=0.99\linewidth]{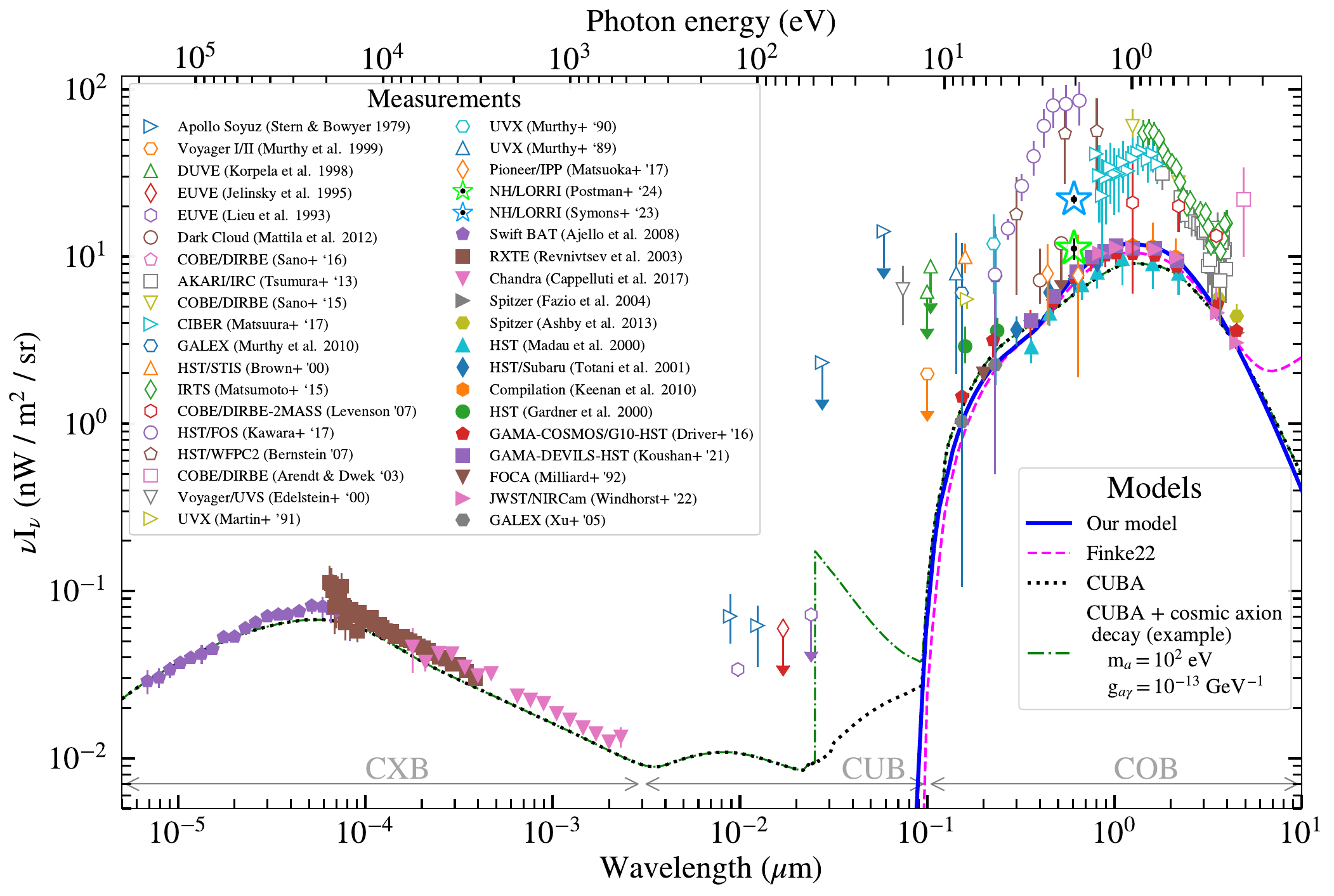}
    \caption{Overview of the CB measurements used here. Data taken from compilations provided by Refs.~\cite{hill_spectrum_2018,Biteau_The_MM_EGAL_spectrum}, in the wavelength range $3 \times 10^{-6}-5\mu$m. We show upper limits (open symbols), lower limits or measurements (filled symbols), and measurements (green and blue star). More details about the different kinds of measurements are given in Sec.~\ref{section:measurements}. Also displayed are the models used here: our model (explained in Sec.~\ref{section:stellarSpectra}), the model developed by \citeauthor{finke_modeling_2022}~\cite{finke_modeling_2022} and the CUBA model~\cite{haardt_radiative_2012}.
    }
    \label{fig:measursOverview}
\end{figure*}

The CB has been measured across the electromagnetic spectrum using various instruments on different satellites. We use the data compilations provided by Refs.~\cite{hill_spectrum_2018,Biteau_The_MM_EGAL_spectrum}, which are shown in Fig.~\ref{fig:measursOverview}.

\subsection{COB}

Following Ref.~\cite{Biteau_The_MM_EGAL_spectrum}, we consider three different types of COB measurements: lower and upper limits, as well as tentative measurements.

Upper limits have been calculated by taking the background flux from optical photometric images~\cite{Kawara_2017,matsumoto_reanalysis_2015,windhorst_jwst_2023,matsuura_new_2017} and subtracting resolved foreground sources such as stars and galaxies, but also diffuse galactic emission.
We treat these values as upper limits, since there might be unaccounted contributions to the images, or the zodiacal light could be wrongly subtracted. These measurements are shown in Fig.~\ref{fig:measursOverview} as open markers.

Lower limits are calculated from galaxy counts~\cite{driver_measurements_2016,windhorst_jwst_2023}. This is done by counting the number and luminosity of galaxies in optical photometric images. Galaxy counts provide the minimum amount of intensity that galaxies would contribute to the CB. Therefore, these measurements can be regarded as lower limits, since any truly diffuse component is not included. They are represented with filled data points in Fig.~\ref{fig:measursOverview}.

The NH data points are shown as stars in Fig.~\ref{fig:measursOverview}. There is a clear mismatch between the value reported in Ref.~\cite{2023lorri} and the rest of the lower limits, which is however much less severe with the latest value reported in Ref.~\cite{2024postmann}. We will study their implications using them either as an upper limit and a measurement of the CB.

\subsection{CUB and CXB}

The CUB intensity is attenuated because of photoionization in the neutral gas of the interstellar medium. Therefore, fewer measurements are generally available in this wavelength regime.
The measurements and upper limits of the CUB we use here have been compiled from spectroscopic surveys. The values shown are the background noise that remains after subtracting all foregrounds from the spectra. Some data were given with a certain confidence interval, defining the background as a value with uncertainties, which we represent in Fig.~\ref{fig:measursOverview} as empty symbols with error bars. Other values are provided as hard upper limits, which are represented with empty symbols with a downward-pointing arrow.

The CXB measurements are derived from x ray surveys by using the background of images after subtraction of individual sources~\cite{cappelluti_chandra_2017}, or using the Earth occultation technique~\cite{ajello_cosmic_2008,revnivtsev_spectrum_2003}. They are depicted as filled symbols in Fig.~\ref{fig:measursOverview}.

\section{Modeling the CB \label{section:theoreticalBackground}}

When a source population produces emission, its contribution to the CB intensity can be calculated as~\cite{kneiske_implications_2002}
\begin{equation} \label{eq:nuInu_general}
    \nu I_{\nu}(\lambda,z) = \frac{c^2}{4\pi\lambda}\int\limits_{z}^{z_\mathrm{max}} \varepsilon_{\nu'}\left(\lambda\frac{1+z}{1+z'}, z'\right)\bigg | \frac{\mathrm{d}t}{\mathrm{d}z'} \bigg | \mathrm{d}z',
\end{equation}
with
\begin{equation}
\hspace{0.3cm}\frac{\mathrm{d}t}{\mathrm{d}z} =\frac{-1}{(1+z) H(z)},
\end{equation}

where we integrate the emissivity $\varepsilon_{\nu}(\lambda, z)$ of the source population over the time it has been emitting. The emissivity depends on wavelength, which is redshifted over cosmic times, and the redshift itself. The maximum redshift $z_\mathrm{max}$ defines the redshift at which the population started emitting.
The Hubble parameter is defined by the $\Lambda$CDM parameters as $H(z) = H_0 \sqrt{\Omega_{\Lambda} + \Omega_\mathrm{m} (1 + z)^3}$.

We proceed by modeling the CB intensity from different sources, such as decaying DM or stars.

\subsection{ALP decay}

We remain agnostic about the exact axion (axions in the following include ALPs) characteristics, but we assume that they are massive enough so that their main interaction is the decay to two photons through $g_{a \gamma}$, which entails a mass $m_a \, \scriptstyle {\gtrsim} $$\,\mathcal{O}(\mathrm{eV})$. 
The decay rate $\Gamma_a$, the inverse of the axion mean life $\tau_{a}$, depends on the axion mass and coupling~\cite{bernal_seeking_2022},
\begin{equation} \label{eq:gammaDefinition}
\begin{split}
    \Gamma_a &= \tau_a ^ {-1} = \frac{\left(m_a c^2\right)^3 g_{a\gamma}^2}{32h} \\
    &= 7.56 \times 10^{-26} \mathrm{s}^{-1} \left(\frac{m_a c^2}{1 \mathrm{eV}}\right)^3 \left(\frac{g_{a\gamma}}{10^{-10} \mathrm{GeV}^{-1}}\right)^2,
\end{split} 
\end{equation}
with \textit{h} being the Plank constant and \textit{c} the speed of light in vacuum.

When an axion decays into two photons, the photons are produced at a wavelength
\begin{equation} \label{eq:decayLambdaRelatMass}
    \lambda_\mathrm{decay} = hc \, \frac{2}{m_a c^2} = 2.48\,\mu\mathrm{m} \left(\frac{m_a c^2}{1 \, \mathrm{eV}}\right)^{-1}\hspace{-0.3cm}.
\end{equation}

\subsubsection{Cosmic axion decay}

The main contribution we study comes from axions distributed throughout the Universe. The emissivity of cosmic axion decay is given by~\cite{overduin_dark_2004}
\begin{equation} \label{eq:axionEmiss}
    \varepsilon_{\lambda, a} (\lambda) = L_\mathrm{h} \delta\left(\lambda - h c \frac{2}{m_a c^2}\right),
\end{equation}
with $L_\mathrm{h}$ the axion luminosity density, which is the luminosity of an axion DM halo of mass $M_\mathrm{h}$ divided by the comoving number of halos $n$,
\begin{equation} \label{eq:luminositydensity}
\begin{split}
    L_\mathrm{h} =& \frac{M_\mathrm{h}}{n} \frac{c^2}{\tau_a} = \Omega_a\rho_\mathrm{crit, 0}\Gamma_a c^2 \\
        =& 1.59 \times 10^{-34} \frac{\mathrm{erg}}{\mathrm{cm^3} \, \mathrm{s}} \left(\frac{m_a c^2}{1 \mathrm{eV}}\right)^3 \left(\frac{g_{a\gamma}}{10^{-10} \mathrm{GeV}^{-1}}\right)^2.
\end{split}
\end{equation}
Assuming the distribution of halos to be homogeneous, we can relate these quantities to the mean axion density of the Universe $\rho_a = \Omega_a\rho_\mathrm{crit, 0}$, with $\Omega_a$ the fraction of DM in the form of axions and $\rho_\mathrm{crit, 0}$ the current critical matter density of the Universe.

The emissivity expression is technically not a Dirac $\delta$ function but a Gaussian distribution, where the width of the decay line is related to the nonrelativistic Doppler effect caused by the velocity dispersion of axions. This width will in general, be different in each system, e.g., in a galactic halo or galaxy cluster.
The broadening is defined as
\begin{equation} \label{eq:broadening}
    \sigma_\mathrm{decay} = 2 \lambda_\mathrm{decay} \frac{v_\mathrm{disp}}{c},
\end{equation}
where $v_\mathrm{disp}$ is the velocity dispersion of the bound axions.
The typical velocity dispersion measured in galaxies is $v_\mathrm{disp} \sim 100 - 300 \, \mathrm{km}/ \mathrm{s}$ \cite{Vdisp1999, Vdisp2017}, which gives a ratio $\sigma_\mathrm{decay} / \lambda_\mathrm{decay} \lesssim 10^{-3}$.
For the contribution to the CB, the ALP decay emissivity is integrated over redshift.
In this case, given the small line width, it is safe to approximate the Gaussian line with a $\delta$ function. More details on this procedure can be found in Ref.~\cite{gong_axion_2016}.

We can plug in the emissivity given by Eq.~\eqref{eq:axionEmiss} in Eq.~\eqref{eq:nuInu_general} and obtain the axion decay contribution to the CB

\begin{equation} \label{eq:axiondecayCOSMIC}
\nu I_{\nu} (\lambda,z) = \frac{\Omega_a\rho_\mathrm{crit, 0} c^4}{64 \pi} \frac{(m_a c^2)^2 g_{a\gamma}^2}{\lambda H(z_{\ast})} \Theta (z_{\ast} - z),
\end{equation}
where 
\begin{equation}
    z_{\ast} = \frac{m_a c^2}{2}\frac{\lambda}{h c}(1+z) - 1. \label{eq:zstar}
\end{equation}
We provide a more detailed derivation in Appendix~\ref{appendix:cosmicdecay}.

\subsubsection{Local halo axion decay}

So far, we have described the contribution that a homogeneous population of halos would give. 
As a complementary emission, we also take into account the DM decay in our local Milky Way halo. For this, we need to integrate the Galactic DM density over the line of sight through the DM halo. For this purpose, we use the software \textsc{clumpy}~\cite{charbonnier_clumpy_2012,bonnivard_clumpy_2016,hutten_clumpy_2019}.
We define the DM column density $S$ as the integral over the DM density along the line of sight (LOS) $\ell(\psi)$, which depends on the LOS angle $\psi(l,b)$ for Galactic longitude and latitude $l,b$,

\begin{equation}
S(\psi) = \int_\mathrm{LOS} \mathrm{d}\ell \, \rho_\mathrm{DM}\left(r(\psi)\right),
\end{equation}
which has units $\mathrm{GeV / cm}^2$.
For our purposes we have chosen a description of our DM halo with a Navarro-Frenk-White (NFW) profile~\cite{1996NFWprofile}
\begin{equation}
    \rho_\mathrm{DM}(r) = \frac{\rho_\odot}{\frac{r}{r_\mathrm{s}}\left(1 + \frac{r}{r_\mathrm{s}}\right)^2},
\end{equation}
with values for the local DM density in the solar neighborhood $\rho_\odot=0.42\, \mathrm{GeV} / \mathrm{cm}^3$, distance from the Sun to the Galactic Center $r_\odot=8.275 \,\mathrm{kpc}$, virial radius $r_\mathrm{vir} = 220 \, \mathrm{kpc}$, and scale radius $r_\mathrm{s}=21 \, \mathrm{kpc}$~\cite{ReviewParticlePhysics, deSalas2020}. 

Since our CB measurements are compilations that encompass many regions of the sky,
we use two different estimates for $S$ that characterize the observations. The smallest possible $S$ is the one that corresponds to a pointing in the Galactic anticenter, which we use as a conservative estimate. A more realistic value can be computed for the NH measurement. It was obtained from observations of different sky regions and observation times~\cite{2023lorri}. We opt for an average value of all pointings, weighted with the observation times.
These two values are reported in Table~\ref{tab:Dfactors}.

\begin{table}[h]
\begin{tabular}{@{}ccc@{}}
\toprule
\midrule
         & \multirow{2}{*}{\begin{tabular}[c]{@{}c@{}}Galactic coordinates \\ $(l,b)$\end{tabular}} & \multirow{2}{*}{\begin{tabular}[c]{@{}c@{}}$S$\\ ($\mathrm{GeV} / \mathrm{cm}^{2}$)\end{tabular}} \\
         &  &   \\ \midrule
Galactic antiCenter & $(0^\circ,\,180^\circ)$ & $1.11 \times 10^{22}$ \\
LORRI observations & Table 2 of~\cite{2023lorri} & $2.21 \times 10^{22}$  \\ \bottomrule
\end{tabular}
\caption{Values of $S$ calculated for two relevant cases discussed in the text. An NFW profile is assumed, with DM halo parameters described in the text.}
\label{tab:Dfactors}
\end{table}

The flux coming from the DM decay is given by~\cite{charbonnier_clumpy_2012}

\begin{align}
        \frac{d\Phi}{dEd\Omega} =& \frac{1}{4\pi} \frac{\Gamma_{a}}{m_a c^2}\frac{dN _\gamma}{dE _\gamma} S \nonumber \\
        =& \frac{1}{4\pi} \frac{\Gamma_{a}}{m_a c^2}\frac{hc}{E^2} \frac{dN _\gamma}{d\lambda} S,
\end{align}
where $dN _\gamma/dE _\gamma = hc/E^2 \, dN _\gamma/d\lambda$ is the spectrum of the axion decay, previously defined as a $\delta$ function.
Here we do not integrate $dN _\gamma/dE _\gamma$ over wavelength, but rather consider its value at specific wavelengths.
Therefore, we cannot use a $\delta$-function approximation as in the cosmic case.\footnote{The value of the flux would be infinite at a certain wavelength.}
So for the local DM decay contribution, we do account for broadening due to velocity dispersion~\cite{overduin_dark_2004}
\begin{equation} \label{eq:deltalongdefinition}
    \frac{dN _\gamma}{d\lambda}\left(\lambda, \lambda_\mathrm{decay}\right) = \frac{1}{\sqrt{2\pi} \sigma_\mathrm{decay} } e^{-\frac{1}{2}\left(\frac{\lambda -\lambda_\mathrm{decay}}{\sigma_\mathrm{decay} }\right)^2},
\end{equation}
Plugging in $\Gamma_{a}$ from Eq.~\eqref{eq:gammaDefinition}, the broadening from Eq.~\eqref{eq:broadening} and the definition of Eq.~\eqref{eq:deltalongdefinition}, we find for the intensity

\begin{eqnarray} 
        \nu I_{\nu}(\lambda) &=& E^2\frac{d\Phi}{dEd\Omega} \nonumber \\
        &=& \frac{c \left(m_a c^2\right)^3}{512\pi~\sqrt{2\pi}~h v_\mathrm{disp}}  g_{a\gamma}^2 \, S\, e^{-\frac{1}{2}\left(\frac{\lambda -\lambda_\mathrm{decay}}{\sigma_\mathrm{decay} }\right)^2} \nonumber \\
        &=& 14.53 \frac{\mathrm{nW}}{\mathrm{sr} \, \mathrm{m^2}} \left(\frac{m_a c^2}{1 \mathrm{eV}}\right)^3\left(\frac{g_{a\gamma}}{10^{-10}\mathrm{GeV}^{-1}}\right)^2 \nonumber \\
        & & \times \left(\frac{S}{1.11 \times 10^{22} \mathrm{GeV} \mathrm{cm}^{-2}}\right)\, \left(\frac{v_\mathrm{disp}}{220 \mathrm{km/s}}\right)^{-1} \nonumber \\
        & & \times e^{-\frac{1}{2}\left(\frac{\lambda -\lambda_\mathrm{decay}}{\sigma_\mathrm{decay} }\right)^2}.\label{eq:axionDecayMW}
\end{eqnarray}

\subsection{Astrophysical sources of the CB}
\subsubsection{Model of COB} \label{section:stellarSpectra}
The excess observed with LORRI could be due to a combination of astrophysical COB emission and ALP decay. 
The publicly available models of the COB usually share only their final results, without providing the means to change the input parameters or add additional sources. Since we aim to study how much additional COB emissivity astrophysical models and additional sources would allow, we develop our own COB model. This allows us to investigate whether it is possible to ease the tension between the COB lower limits and the LORRI measurements and to investigate potential degeneracies between the ALP and astrophysical contributions. As we will see below, the stellar COB contribution is completely fixed by the observations of galaxy emissivities, erasing any such degeneracy.

The COB is dominated by stellar emission, which is partly absorbed by interstellar dust. Other known astrophysical components include AGN emission which, however, dominates in the ultraviolet and accounts for at most 10\% of the COB~\cite{2011Dominguez_agn, finke_modeling_2022}, or intrahalo light, whose contribution is expected to be negligible ~\cite{cheng_probing_2021,mitchell-wynne_ultraviolet_2015,bernal_seeking_2022}.

For our COB model, we follow the methodology of Ref.~\cite{finke_modeling_2022}, hereafter \citetalias{finke_modeling_2022}. Synthetic stellar spectra are calculated with the \textsc{Starburst99} code~\cite{leitherer_starburst99_1999,leitherer_effects_2014}, using a Kroupa initial mass function (IMF) with $M_\mathrm{min}=0.1\,\mathrm{M}_\odot$ and $M_\mathrm{max}=100\,\mathrm{M}_\odot$~\cite{kroupa2002}. We maintain the dust and star formation rate parametrizations used in Ref.~\cite{finke_modeling_2022}, but parametrize the metallicity evolution following Ref.~\cite{2022Tanikawa}.
We fit our model to various datasets of COB intensity, stellar emissivities, star formation rate, and metallicities \cite{Biteau_The_MM_EGAL_spectrum, hill_spectrum_2018, the_fermi-lat_collaboration_gamma-ray_2018, finke_modeling_2022, 2020Zmeasurs}, obtaining a reduced $\chi^2$ value over degrees of freedom of $\chi^2 / \mathrm{d.o.f.} = 2.7$.
Adding these datasets to the COB fit essentially fixes the stellar contribution, as the introduction of the ALP decay to the CB does not affect the metallicity, star formation rate, and galaxy emissivities. 
As a consequence, we choose to fix the COB contribution to its best fit and add additional contributions such as ALP decay on top of it. 
We also tested adding other components (intrahalo light, stripped stars) to check whether their contributions could help explain the LORRI excess.
However, these other contributions are not bright enough and we can rule them out as possible sources for the excess.

This model is shown in Fig.~\ref{fig:measursOverview}. Further details about this model and tests with different contributions are detailed in Appendix~\ref{appendix:cob_charact}.

We use two preexisting COB models to compare with ours. 
The first one comes from~\citetalias{finke_modeling_2022}, which encompasses the COB and CIB and models the contributions of stars and dust.
The second one is the model of Ref.~\cite{haardt_radiative_2012}, hereafter \citetalias{haardt_radiative_2012}, which covers the frequency range from the CXB to the COB and models all the dominant processes of radiative transfer. They are both shown in Fig.~\ref{fig:measursOverview}.
In the central region of the COB, the model with the smallest intensity is \citetalias{haardt_radiative_2012}; while at the smaller COB wavelengths it is \citetalias{finke_modeling_2022}. These behaviors are also shown in the top part of Fig.~\ref{fig:cobAndParams}.

\subsubsection{Model of CUB and CXB}

The main components of the CUB are emission from young stars and interstellar nebulae, scattered by gas and hot intercluster gas~\cite{hill_spectrum_2018}. 
The efficiency of neutral hydrogen of absorbing ultraviolet light is almost unity~\cite{hill_spectrum_2018}, which results in a very low intensity of the CUB compared to the COB, see Fig.~\ref{fig:measursOverview}.

The CXB is mainly produced by accretion disks around AGN, which create high-energy photons via thermal Bremsstrahlung effect~\cite{hill_spectrum_2018}.

In contrast to the COB, we solely rely on the \citetalias{haardt_radiative_2012} predictions for the CUB and CXB.

\section{Data analysis} 
To constrain the axion parameters, we treat the (putative) direct detections of the CB as one-sided upper limits. Under the assumption that the likelihoods for the data points follow Gaussian statistics, we arrive at
\begin{equation} \label{eq:likelihoodUpperLims}
    \begin{split}
    \chi^2 = \sum_i &\left(\frac{\nu I_{\nu_\mathrm{model}} (\lambda_i) - \nu I_{\nu_i} }{\sigma_i}\right)^2 \\
    &\times \, \Theta \left(\nu I_{\nu_\mathrm{model}} (\lambda_i) - \nu I_{\nu_i}\right),
    \end{split}
\end{equation}
where $\Theta(x)$ is the Heaviside step function, and $\nu I_{\nu_\mathrm{model}}$ is the sum of the CB and axion decay contribution. The CB contribution is fixed by either the predictions of the \citetalias{haardt_radiative_2012} or \citetalias{finke_modeling_2022} models, or the predictions from our own model. The latter is, in turn, fixed by the fit to various datasets as described in Appendix~\ref{appendix:cob_charact}.

For the COB upper limits and CUB putative measurements we take the value of $\nu I_{\nu_\mathrm{model}}$ as the value given by the model at the given wavelength $\lambda_i$, but for the CXB measurements we calculate the mean value of $\nu I_{\nu}$ over each energy bin.

Alternatively, we treat the NH data point as a measurement. 
In this case, we take the LORRI quantum efficiency $P(\lambda)$ into account. The wavelength dependence of $P(\lambda)$ is given in Fig.~22 of Ref.~\cite{2020lorri_specs}. This is necessary as the spectral shape of the axion decay signal is not flat, as was assumed for the measurement of the COB. The mean flux measured over the detector is given by

\begin{equation} \label{eq:meanFlux}
    I_{\nu, \mathrm{mean}}  = \frac{\int I_\nu(\lambda) \, P (\lambda) \, d\lambda/\lambda}{\int P(\lambda) \, d\lambda/\lambda}.
\end{equation}
The term $I_\nu (\lambda)$ is the spectral intensity measured with LORRI, which in our case is the sum of astrophysical COB and axion decay. This expression is taken from Ref.~\cite{1986HiA.....7..833K}, but we have used intensities in the calculations instead of flux.
Intensity $I_\nu$ and flux $f_\nu$ are related by the integration over solid angle and since $I_\nu$ is isotropic, it does not matter if we use intensity or flux.
Furthermore, the mean wavelength associated with this flux measurement is given by~\cite{1986HiA.....7..833K}
\begin{equation} \label{eq:meanWavelength}
    \langle\lambda\rangle = \frac{\int P(\lambda) \, I_\nu (\lambda) \, d\lambda}{\int P(\lambda) \, I_\nu (\lambda) \, d\lambda/\lambda}.
\end{equation}
With axion decay present in the wavelength range of the bandpass, both the mean intensity and wavelength depend on $m_a$ and $g_{a\gamma}$.

With these considerations, we can now compare our model predictions with the LORRI measurement through
\begin{equation} \label{eq:likelihoodNH}
    \chi^2 = \left(\frac{\nu I_{\nu_\mathrm{mean, model}} (\langle\lambda\rangle) - \nu I_{\nu_{,\mathrm{NH}}} \times {\lambda_\mathrm{pivot}}/{\langle\lambda\rangle} }{\sigma_\mathrm{NH}}\right)^2,
\end{equation}
where $\lambda_\mathrm{pivot} = 0.6076\,\mu$m is the pivot wavelength of the LORRI bandpass associated with the LORRI measurement~\cite{2020lorri_specs}. The ratio ${\lambda_\mathrm{pivot}}/{\langle\lambda\rangle}$ comes from shifting the NH measurement to the same mean wavelength as our calculation, so we can perform a consistent comparison.

\section{Results} \label{section:results}

We first discuss the results for cosmic axion decay, followed by the results for axion decay in the local halo, and finally the results for the LORRI measurement.

\subsection{Cosmic axions}
We add the cosmic axion component given in Eq.~\eqref{eq:axiondecayCOSMIC} to the COB models and explore the axion-photon coupling parameter space for which the axion contribution does not overshoot the CB upper limits.
An example of axion decay added to the \citetalias{haardt_radiative_2012} model, with arbitrary values for the axion parameters, can be seen in Fig.~\ref{fig:measursOverview}.

\begin{figure}
    \centering
    \includegraphics[width=0.99\linewidth]{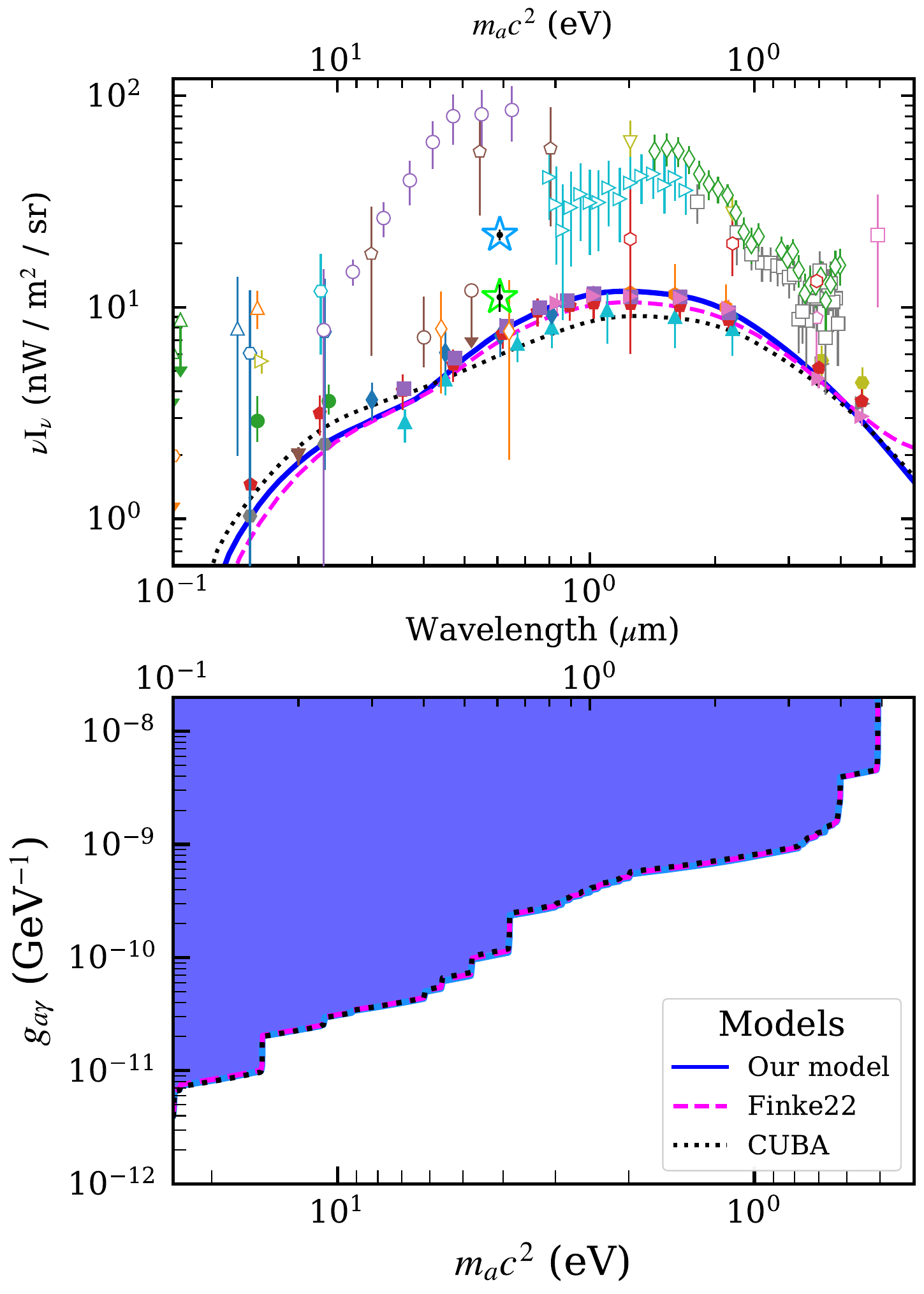}
    \caption{Top: COB measurements and models used or calculated in this work. Bottom: axion parameter space that is constrained by the COB upper limits by considering the cosmic axion contribution (for each of the discussed COB models). The excluded region (at a 2$\sigma$ confidence level) is above the different lines, shaded in blue. Note the descending $x$ axis, following the correlation between axion mass and photon wavelength in Eq.~\eqref{eq:decayLambdaRelatMass}.
    }
    \label{fig:cobAndParams}
\end{figure}

Treating the NH data point as an upper limit, along with the other upper limits, we follow the methodology in Eq.~\eqref{eq:likelihoodUpperLims} and calculate exclusions at the $2\sigma$ confidence level. 
In Fig.~\ref{fig:cobAndParams} we show our constraints derived from COB data (bottom panel) and an enlargement of the COB measurements themselves (top panel). 
In the upper (lower) panel, the upper (lower) $x$ axis shows the axion mass, which is  related to wavelength through Eq.~\eqref{eq:decayLambdaRelatMass}.
One can observe a correspondence between the constraints and the COB data points: for upper limits of comparatively low intensity (such as the AKARI/IRC~\cite{Tsumura2013_akari} or the Dark Cloud measurements~\cite{Mattila2012_darkcloud}), the limits improve toward higher axion masses and result in sharp features.
We have used the LORRI measurement of \cite{2023lorri} for the calculations, but this value has little relevance, since the constraints in that region are dominated by the Dark Cloud measurements.
There is a sharp cut in the constraints at higher wavelengths since the parameter space we can scan with this methodology ends at the position of the highest wavelength data point we consider.
Note that lower decay wavelengths correspond to higher axion masses, and the tails in the decay spectra of these axions (see Fig.~\ref{fig:measursOverview} for an example) also contribute to the $\chi^2$. 
The different COB models result in very similar axion constraints.

\begin{figure*}[ht!]
    \centering
    \includegraphics[width=0.99\linewidth]{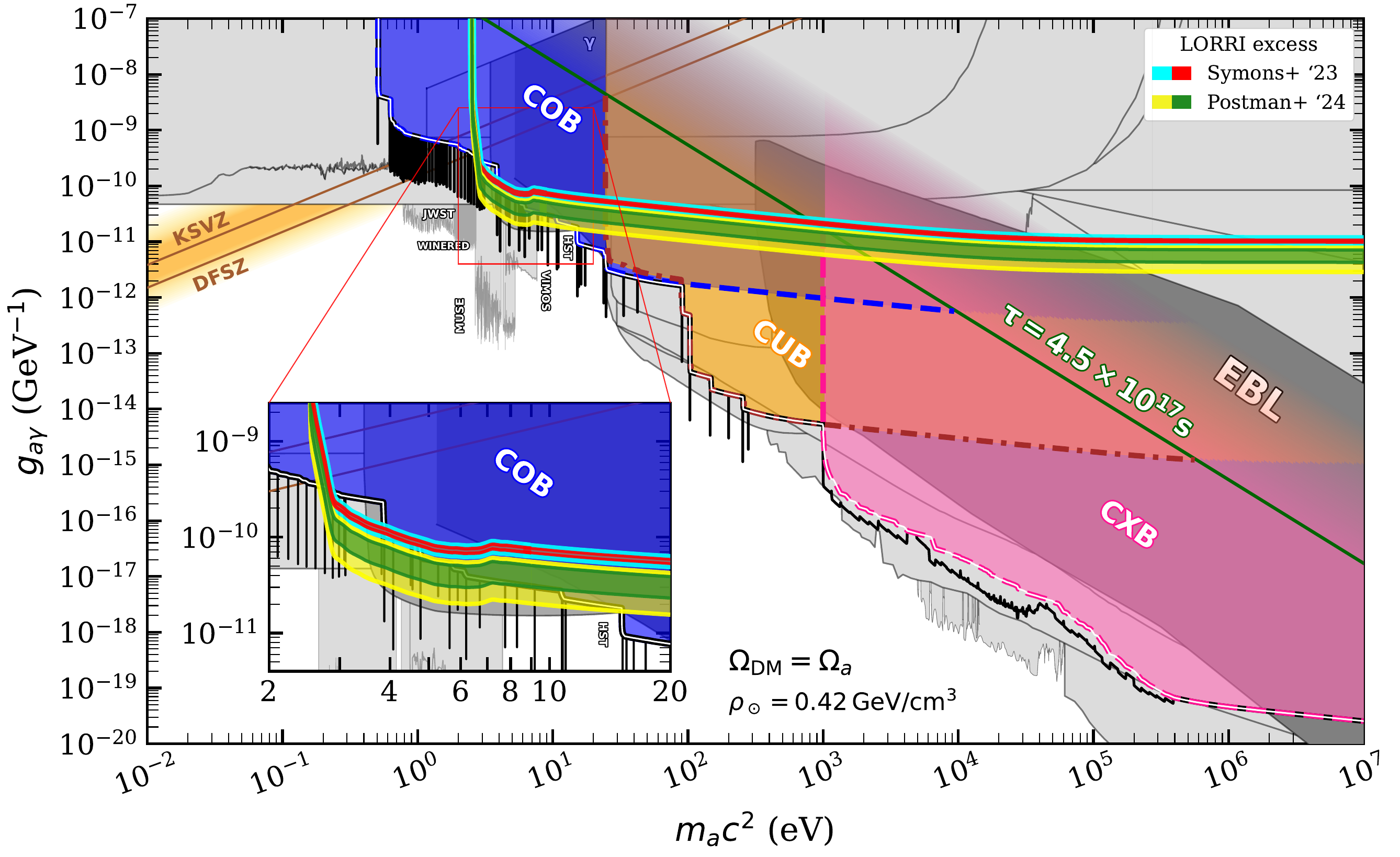}
    \caption{Our and previous constraints on the axion and ALP parameter space, using the \citetalias{haardt_radiative_2012} model. 
    Confidence intervals at 1$\sigma$ (2$\sigma$) that could explain the NH excess are shown in red (blue) for the values of Ref.~\cite{2023lorri} and in green (yellow) for the values of Ref.~\cite{2024postmann}.
    Constraints at 2$\sigma$ confidence using the NH measurement from \cite{2023lorri} as an upper limit, together with the rest of upper limits from Fig.~\ref{fig:measursOverview}, are shown as shaded blue, yellow, and pink regions for the COB, CUB, and CXB, respectively. The inset shows the comparison between COB models in Fig.~\ref{fig:measursOverview}. 
    The black line shows the overall constraint when also the axion decay in the local DM halo is accounted for.
    Limits from previous analyses are depicted in gray and are taken from the compilation of Ref.~\cite{AxionLimits}.
    }
    \label{fig:paramsGITHUB}
\end{figure*}

In Fig.~\ref{fig:paramsGITHUB} we show the constraints of the axion-photon coupling for the full wavelength range we consider,  again using the methodology of Eq.~\eqref{eq:likelihoodUpperLims}.
We have calculated this set of constraints using the \citetalias{haardt_radiative_2012} model as the astrophysical contribution, adding the cosmic axion decay on top.
Using only the data points from the COB, CUB, and the CXB, we arrive at the constraints shown as filled blue, yellow, and pink regions, respectively. These constraints are at $2\sigma$ confidence. 
The constraint from the whole spectral region, using all the data points, is shown as the white line.
We guide the figure with a green straight line labeled $\tau = 4.5 \times 10^{17}$s, which links the axion parameters following Eq.~\eqref{eq:gammaDefinition} using the age of the Universe~\cite{2020plank} as the mean axion lifetime.
For all the filled regions we can observe a power-law behavior of the constraints toward higher masses. This is due to the definition of our $\chi^2$ function and further detailed in Appendix~\ref{appendix:powerLaws}.

\subsection{Decay in the Galactic halo}

Next, we consider the axion decay contribution from the local Milky Way halo, given by Eq.~\eqref{eq:axionDecayMW}.

We start with the addition of the host axion decay in the wavelength bands of the COB and CUB. The observations at these wavelengths have been carried out on a large variety of instruments with different bandpasses using different data analysis techniques. For simplicity, we opt to use the measured intensities at face value and at the reported wavelengths, i.e., we neglect the finite bandpass, which we only consider in the case of the LORRI measurements. Incorporating the specific instrumental responses is left for future study. 

\begin{figure}
    \centering
    \includegraphics[width=0.99\linewidth]{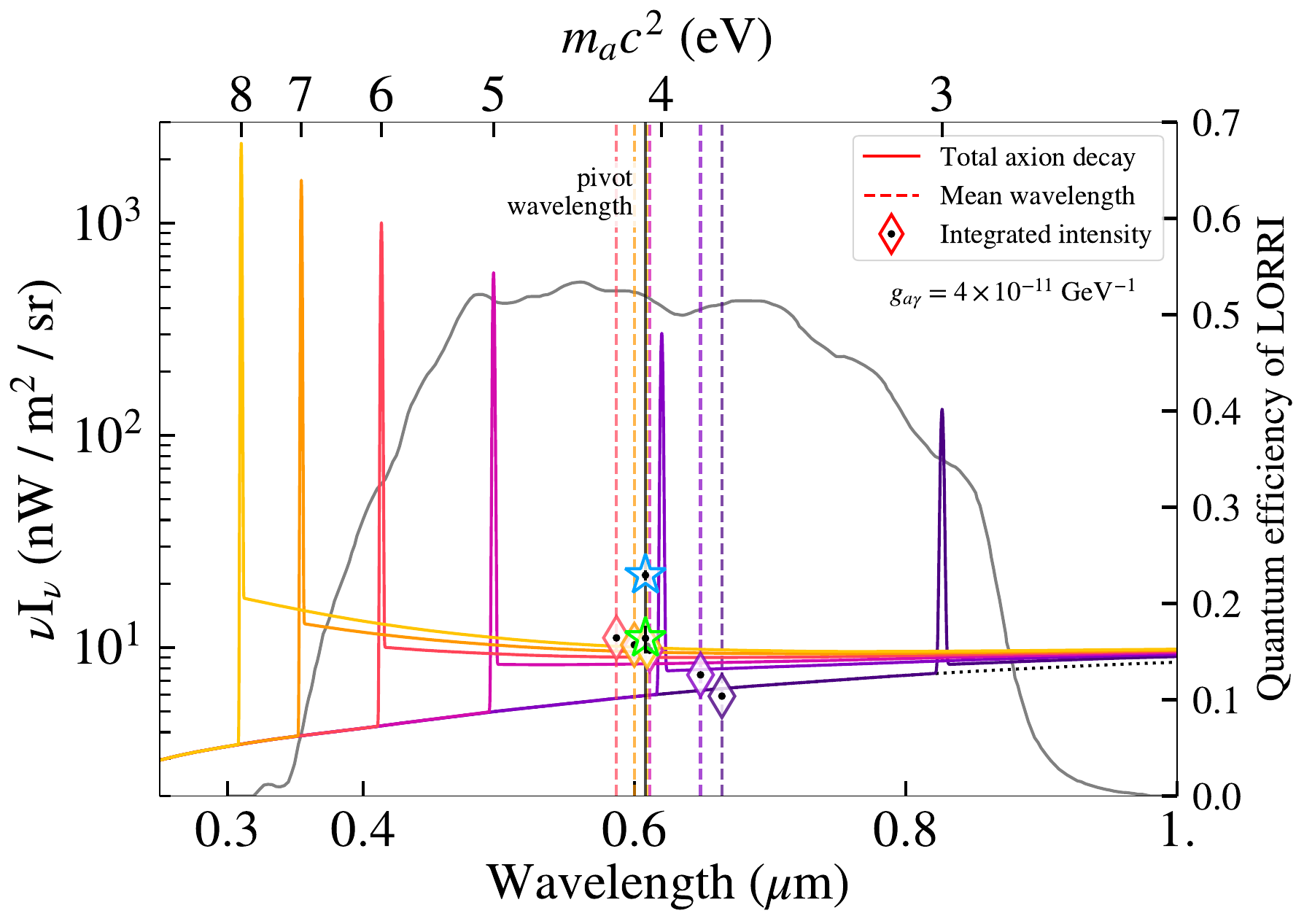}
    \caption{Cosmic and local axion decay spectra for axions at different masses.
    The mean wavelength of each spectrum is shown as a vertical line sharing the color of each axion spectrum calculated with Eq.~\eqref{eq:meanWavelength}, as well as the mean flux from Eq.~\eqref{eq:meanFlux}. The dotted line is the \citetalias{haardt_radiative_2012} spectrum. The gray line and the right $y$ axis give the quantum efficiency of the LORRI detector.
    }
    \label{fig:axiondecayQEpretty}
\end{figure}

In Fig.~\ref{fig:axiondecayQEpretty} we show examples of the decay of an axion with masses between $m_a=3$ and $8\,$eV including both cosmic and host contributions on top of the \citetalias{haardt_radiative_2012} model. 
The coupling constant is the same for all considered masses.
We can see that the host contribution is only relevant at wavelengths corresponding to energies at half of the axion rest mass, whereas the cosmic decay contribution dominates the decay spectrum at longer wavelengths. Therefore, the Galactic halo decay mostly affects the constraints close to the corresponding wavelengths at which we have measurements.

We consider the single data points one at a time. For each data point, we calculate the axion mass that would contribute to the intensity following Eq.~\eqref{eq:decayLambdaRelatMass} and then the $\nu I_{\nu_\mathrm{model}}$ necessary to overshoot the 2\,$\sigma$ uncertainty using Eq.~\eqref{eq:likelihoodUpperLims}. We determine the corresponding coupling constant following Eqs.~\eqref{eq:axiondecayCOSMIC} and~\eqref{eq:axionDecayMW}.
We use the column density $S$ from the Galactic anticenter, yielding conservative constraints.
The only exception is the LORRI observation, for which we use the values for $S$ listed in Table~\ref{tab:Dfactors}.
The result is plotted in Fig.~\ref{fig:paramsGITHUB} as the black line, where we can see sudden dips departing from the white line in the COB and CUB at axion masses where we have a data point at the correct wavelength.

In the CXB region, we perform a similar analysis, but we include the widths in wavelength of the spectral data points. Instead of assuming a monochromatic decay and only checking the flux on the associated wavelength, we calculate the mean value of the intensity of the axion decay in each bin and use this value for our model estimate $\nu I_{\nu_\mathrm{model}}$. 
We assume a constant exposure in each wavelength bin. 
This results in constraints in the CXB that are not as sharp as in the COB and CUB, see the black line in Fig.~\ref{fig:measursOverview}.

\subsection{Analysis of LORRI observations}
\begin{figure}
    \centering
    \includegraphics[width=0.99\linewidth]{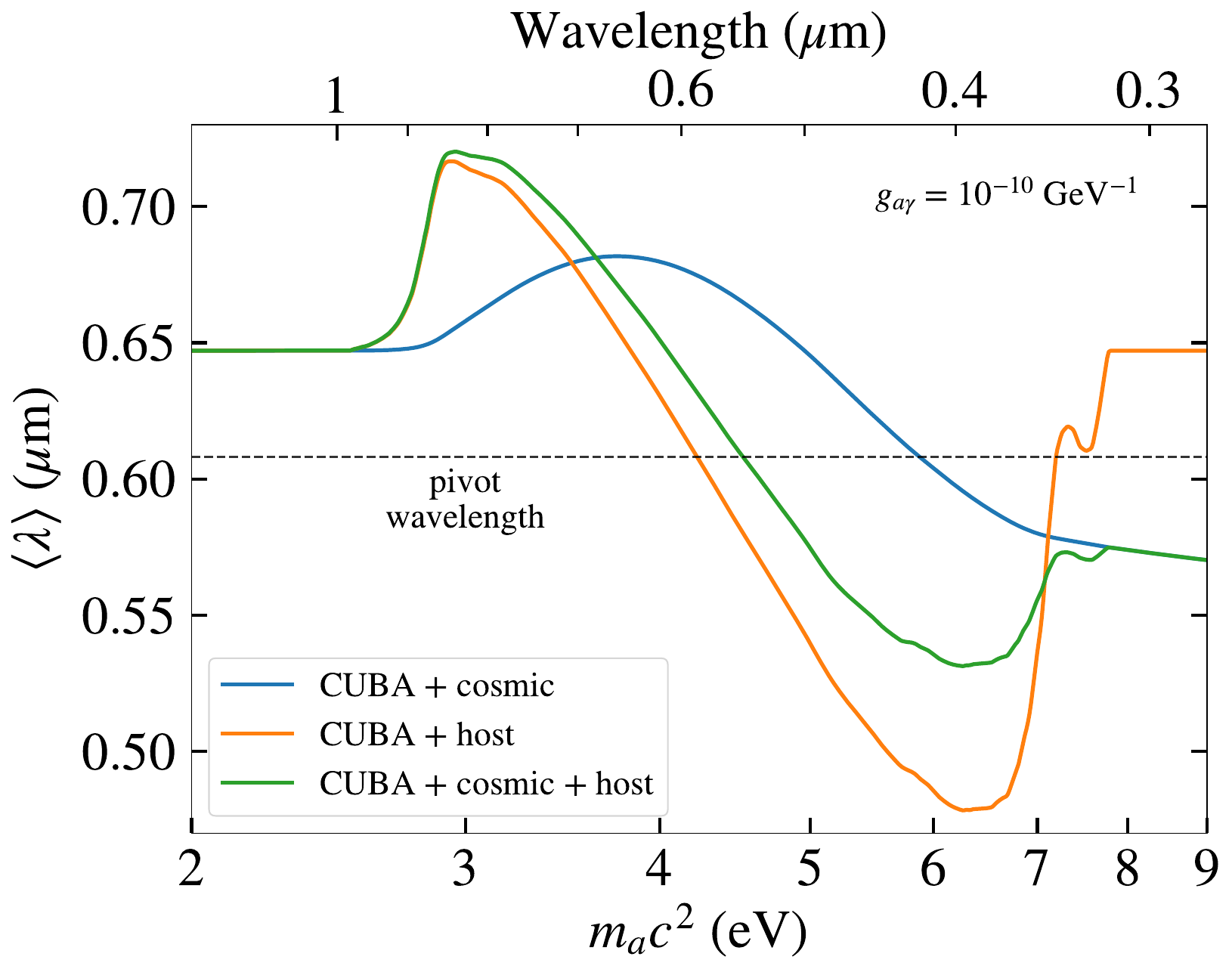}
    \caption{Mean wavelength associated with axion decays, as a function of axion mass following Eq.~\eqref{eq:meanWavelength} assuming constant coupling constant.
    }
    \label{fig:meanLambda}
\end{figure}

Finally, we turn our attention to the LORRI measurement. In this analysis, we make use of the quantum efficiency and finite bandpass of the detector and find the axion parameters for which the \citetalias{haardt_radiative_2012} model and axion contributions fit the NH data point. 
To this end, we first calculate the mean flux and wavelength of the axion decay spectrum following Eqs.~\eqref{eq:meanFlux}~and~\eqref{eq:meanWavelength}, respectively.
The quantum efficiency of the LORRI instrument is shown in Fig.~\ref{fig:axiondecayQEpretty} as a gray line (see the right $y$ axis). 
The pivot wavelength of the detector is shown as the vertical black line.
As mentioned before, in this figure, we also show examples of axion decay spectra for different masses.
For these spectra we calculate their associated mean wavelength with Eq.~\eqref{eq:meanWavelength} (dotted vertical lines) and the corresponding mean flux with Eq.~\eqref{eq:meanFlux} (black dots and colored diamonds).
The spectrum, mean wavelength, and flux for each axion mass share the same color.
We can see that the mean wavelength varies with mass and we show this dependency in  Fig.~\ref{fig:meanLambda} for a fixed photon-axion coupling and the \citetalias{haardt_radiative_2012} model. 
Compared to the pivot wavelength (gray dashed line), which corresponds to a flat spectrum, we see several features in the mean wavelengths, in particular when the host decay is included (orange and green lines). 
These features stem from the sharply peaked decay spectrum folded with the LORRI quantum efficiency. 
At high and low masses, when the decay peak is not within the LORRI passband anymore, we still observe a shift of the mean wavelength compared to the pivot wavelength, which is due to the \citetalias{haardt_radiative_2012} model itself. 

Having the mean wavelength and flux, we calculate the $\chi^2$ value from Eq.~\eqref{eq:likelihoodNH}.
In Fig.~\ref{fig:paramsGITHUB}, we show the $1\,\sigma$ and $2\,\sigma$ confidence intervals (corresponding to a $\Delta\chi^2 = 2.30, 5.99$ for 2 degrees of freedom) of the axion parameters that would explain the LORRI measurements of Ref.~\cite{2023lorri} (Ref. \cite{2024postmann}) in red and blue (green and yellow).
The preferred coupling constant lies between $10^{-10}$ and $10^{-11}\,\mathrm{GeV}^{-1}$ for masses above $m_a \gtrsim 2\,$eV.
The preferred regions for the two LORRI measurements have the same shape, the difference between them being the strength of coupling necessary to obtain he excess value.

In the inset of Fig.~\ref{fig:paramsGITHUB}, we show an enlarged view of the axion constraints in the COB regime.
In this region, also the DM host decay contributes to the LORRI measurement up to roughly $m_a\sim 7\,$eV, which can be seen in the shape of the confidence intervals. 
In this region, up to $m_a\sim 10\,$eV, axions could both explain the NH excess and not violate the other CB upper limits.
However, this region has been excluded by other experiments such as the observations of globular clusters~\cite{Dolan2022}, spectroscopic searches with the Multi Unit Spectroscopic Explorer (MUSE)~\cite{todarello_robust_2023} or study of COB anisotropies with the Hubble Space Telescope (HST)~\cite{carenza_probing_2023}.

\section{Discussion} \label{section:discusssion}

We find that the different COB models provide very similar constraints in the axion parameter space, as we see in Fig.~\ref{fig:cobAndParams}. 
Thanks to the implementation of our own COB model, detailed in Appendix~\ref{appendix:cob_charact}, we test different characterizations of the stellar population, such as stellar spectra or dust descriptions. 
Our best-fit model still has a reduced $\chi^2$ value larger than unity, its largest contribution coming from the fit to stellar emissivities.
We find that the best-fit COB model is completely fixed by the observational data, leaving no room for altering its spectrum significantly to potentially relieve tension with the LORRI data point.
We also find that other astrophysical contributions to the COB such as AGN, intrahalo light, or a population of stars with stripped envelopes are subdominant, see Appendix~\ref{appendix:additional}.
Looking at Fig.~\ref{fig:paramsGITHUB}, we see that only a small region of the axion parameter space, located at $m_a \sim 4-5\,$eV, could explain the NH data point, while simultaneously not overshooting other CB upper limits. As mentioned before, this parameter space is ruled out by other experiments such as globular clusters~\cite{Dolan2022} or MUSE~\cite{todarello_robust_2023}.

Previous limits that use a similar technique as ours are shown in dark gray in Fig.~\ref{fig:paramsGITHUB}.
In particular, Ref.~\cite{cadamuro_cosmological_2012}, labeled ``EBL'' in the figure, also used the CB spectrum available at that time to constrain axion parameters. 
Our constraints improve their results by 1 order of magnitude over a wide range of axion masses, also extending it to lower masses. We even rule out the Dine-Fischler-Srednicki-Zhitnitsky (DFSZ) and Kim-Vainshtein-Shifman-Zakhar (KVSZ) axion models in between $m_a = 2$ and $30 \, \mathrm{eV}$. Our results are more restrictive because we include the latest CB measurements and the astrophysical contribution to the CB.
Limits labeled HST~\cite{nakayama_anisotropic_2022} and $\gamma$~\cite{bernal_seeking_2022} are based on the NH measurement as well but different methodologies were used for their derivation.
For the HST constraint, it was studied how the LORRI excess impacts COB anisotropy. The $\gamma$ exclusion used the axion contribution to the COB in the resulting increase to the optical depth for very high-energy $\gamma$~rays. 
Our results yield, in general, similar constraints compared to these studies.
We extend the analysis by Ref.~\cite{bernal_cosmic_2022} by including additional CB measurements in our analysis.
As a consequence, only a small region of the axion parameter space is allowed to explain the LORRI data point.

Further constraints in the $m_a\sim1-10$ eV regime, such as MUSE~\cite{todarello_robust_2023}, Visible Multiobject Spectrograph~\cite{grin_telescope_2007}, and James Webb Space Telescope (JWST)~\cite{janish_hunting_2023} as labeled in Fig.~\ref{fig:paramsGITHUB}, stem from spectroscopic measurements. This method proves to be more constraining compared to our analysis, while also being more restricted by the detection range of instruments.

\section{Summary and conclusions} \label{section:conclusions}
In this work we have investigated the impact of axion or ALP decay on the CB, specifically from optical wavelengths to x-ray energies.
We have put particular emphasis on the optical regime due to the tentative excess observed with LORRI~\cite{2023lorri}, on board the NH probe, although latest measurements suggest it is not significant anymore~\cite{2024postmann}. Such an excess would be challenging to explain with known astrophysical sources.

To this end, we have implemented our own COB model. We have tested various model assumptions by fitting them to observational data and have found our best-fit model, which we show in the main text.
We compare our results with those of \citetalias{finke_modeling_2022} and \citetalias{haardt_radiative_2012} and find that different COB models provide similar results. Specifically, when applying these models to the axion decay, the constraints are very similar.

Concerning the axion-photon parameter space, we find only a small region of the parameter space around $4-5\,\mu$m can explain the NH point as a direct measurement of the CB while not overshooting the CB upper limits.
This parameter space has been ruled out by other experiments.
We also find that this mild excess is unlikely caused by either intrahalo light or stripped stars. As we show in Appendix~\ref{appendix:additional}, intrahalo light peaks at wavelengths above the LORRI bandpass, and stripped stars contribute around 1\% to the COB.
However, we note again that the significance of the excess is only at $1.4\,\sigma$ above the COB level measured with galaxy counts. 
The parameter space that explains the excess reported in Ref.~\cite{2023lorri} is fully constrained by other experiments.
We have also calculated new constraints in the axion-photon parameter space assuming the NH observation as another upper limit, and combined its analysis to a compilation of CB measurements. Our constraints improve the previous ones from  Ref.~\cite{cadamuro_cosmological_2012} on average by a factor of 10 (see Fig.~\ref{fig:paramsGITHUB}). 
Assuming $\Omega_a = \Omega_\mathrm{DM}$, we rule out the DFSZ and KVSZ axion models in between $m_a = 2$ and $30 \, \mathrm{eV}$ with the decay of cosmic axions.

Axion decay in the Galactic halo would create narrow peaks of flux in the COB and CUB. This leads to strong constraints at axion masses corresponding to a wavelength where CB measurements are available. In the CXB, the effect is less striking, as we have averaged the signal over each energy bin (see Fig.~\ref{fig:paramsGITHUB}). 
For the constraints we conservatively assumed the DM column density $S$ of the Galactic anticenter, which is insensitive to the adopted DM profile and scales linearly with the assumed DM density in the Solar System (we used $\rho_\odot = 0.42\,$GeV/cm$^3$). There is one exception to this: for the NH measurements, we calculated the weighted mean of $S$, based on the observation times and pointings of the probe.

As an outlook, new measurements of the CB, i.e. with the JWST, would improve our predictions: both direct and indirect measurements of the CB, as well as better characterization of stellar parameters.
Moreover, further sources could contribute to the CB. For example, a change in the stellar formation rate at high redshift due to primordial black holes~\cite{cappelluti_exploring_2022} or the existence of Population III stars~\cite{2009Raue_pop3} would increase the astrophysical CB component. Another source could be dark stars~\cite{2012Maurer_darkStars}.
The decay in the local halo is another sensitive probe. Detailed analysis with correct values of the DM column density $S$ for each observation could be promising, which we aim to explore in a future publication. 
Furthermore, we plan to extend our analysis to higher energies and include the extragalactic diffuse $\gamma$-ray background, and lower energies to account for the infrared background.

\begin{acknowledgments}
The authors would like to thank Ciaran O'Hare for his thoughtful comments and discussions. We would also like to thank Todd Lauer and Marc Postman for sharing the preliminary updated LORRI measurement. 
We further thank Marco Ajello and Kari Helgason for providing us with the data of the SFR in machine-readable format. 
The authors acknowledge support from the European Research Council (ERC) under the European Union’s Horizon2020 research and innovation program Grant Agreement No. 948689 (AxionDM) and from the Deutsche Forschungsgemeinschaft (DFG, German Research Foundation) under Germany’s Excellence Strategy – EXC 2121 “Quantum Universe”– 390833306. This article is based upon work from COST Action COSMIC WISPers CA21106, supported by COST (European Cooperation in Science and Technology).
\end{acknowledgments}

\bibliography{MyLibrary} 

\appendix

\section{Derivation of cosmic axion decay from first principles} \label{appendix:cosmicdecay}

The emissivity of axion decay is given by~\cite{overduin_dark_2004}

\begin{equation}
    \varepsilon_{\lambda, a} (\lambda) = L_\mathrm{h} \delta\left(\lambda - h c \frac{m_a c^2}{2}\right),
\end{equation}
from which we see
\begin{equation}
    \varepsilon_{\nu, a} (\lambda) = L_\mathrm{h} \frac{\lambda^2}{c} \delta\left(\lambda - h c \frac{m_a c^2}{2}\right).
\end{equation}
Inserting this into Eq.~\eqref{eq:nuInu_general}, we obtain
\begin{align}
    \nu I_{\nu}(\lambda,z) =& \frac{c \lambda}{4\pi} L_h (1+z)^2 \\
    & \cdot \int^{z_{max}}_{z_0} \frac{dz'}{(1+z')^3 H(z')} \, \delta\left(\lambda\frac{1+z}{1+z'} - \frac{2 h c}{m_ac^2}\right). \nonumber
\end{align}
Changing variables of the integration from $z'$ to $y = \lambda\frac{1+z}{1+z'} - \frac{2 h}{m_ac}$, we define $z_\star$ so that $y(z_\star)=0$. Now we can easily perform the integral and find
\begin{equation}
    \nu I_{\nu}(\lambda,z) = \frac{c}{4\pi} L_\mathrm{h} \frac{\Theta (z_{\ast} - z)}{H(z_{\ast})} \frac{h c}{\lambda}\frac{2}{m_a c^2}.
\end{equation}
Using the definition of $L_\mathrm{h}$ given by Eq.~\eqref{eq:luminositydensity}, we obtain Eq.~\eqref{eq:axiondecayCOSMIC}.

\section{COB model} \label{appendix:cob_charact}
Here, we provide details of the COB modeling.\footnote{The model is available online at \url{https://github.com/axion-alp-dm/EBL_calculation}}
First, we describe the complementary observational data and the theoretical background of the model. This leads to the fitting procedure and results. We also discuss how different model assumptions affect our results, as well as the impact additional subdominant components such as intrahalo light and stripped stars.

\subsection{Emissivities, star formation rate, and metallicity evolution}

For our COB model, we make use of additional observational data, which we describe in this subsection.

First, the galaxy emissivities have been obtained in the COB region. These measurements come from observations of resolved galaxies, detailing how much light these galaxies emit per wavelength and redshift. A compilation of emissivity values can be found in \citetalias{finke_modeling_2022}, and are plotted in Fig.~\ref{fig:emissivities}.

Data of the star formation rate $\rho_{\star}$ as a function of redshift has been compiled in Ref.~\cite{the_fermi-lat_collaboration_gamma-ray_2018}. It is defined as the quantity of stars that form at any epoch per volume. There are different methods to measure $\rho_{\star}$, such as luminosity measurements of galaxies in the ultraviolet and infrared, or using the nebular line emission that characterizes star formation~\cite{madau_cosmic_2014}. They are plotted in Fig.~\ref{fig:sfr}.

Finally, we use measurements of the mean metallicity of the Universe as a function of redshift, in order to determine the metallicity of the stellar populations at any point in time. The observational data are taken from~\cite{2020Zmeasurs}. They are plotted in Fig.~\ref{fig:metallEvol}.

\subsection{Details on the theoretical modeling of the COB}

\subsubsection{Stellar emissivities and synthetic spectra}

A simple stellar population (SSP) is defined as a group of stars born from the same gas cloud, which then evolve with no further star formation~\cite{leitherer_starburst99_1999}. The birth of the SSP lasts for megayears, so for cosmological purposes it can be assumed to be instantaneous. The stars in an SSP are assumed to share the same metallicity and age.

The emissivity of the stellar population can be calculated as
\begin{equation} \label{eq:emissStellar}
\begin{split}
    \varepsilon_{\nu_\mathrm{stellar}}(\lambda, z)& = f_\mathrm{esc, dust} \\
    \times& \int^{z_\mathrm{max}}_{z} L_{\nu}(t(z) - t(z'))\,\rho_{\star}(z')\bigg|\frac{dt'}{dz'}\bigg|dz',
\end{split}
\end{equation}
where $L_{\nu}$ is the luminosity of the SSP, $\rho_{\star}$ is the stellar formation rate, and $f_\mathrm{esc, dust}$ is the fraction of light that escapes dust absorption.
We integrate up to redshift $z_\mathrm{max} = 40$.
The individual terms will be discussed below.

\begin{figure}
    \centering
    \includegraphics[width=0.9\linewidth]{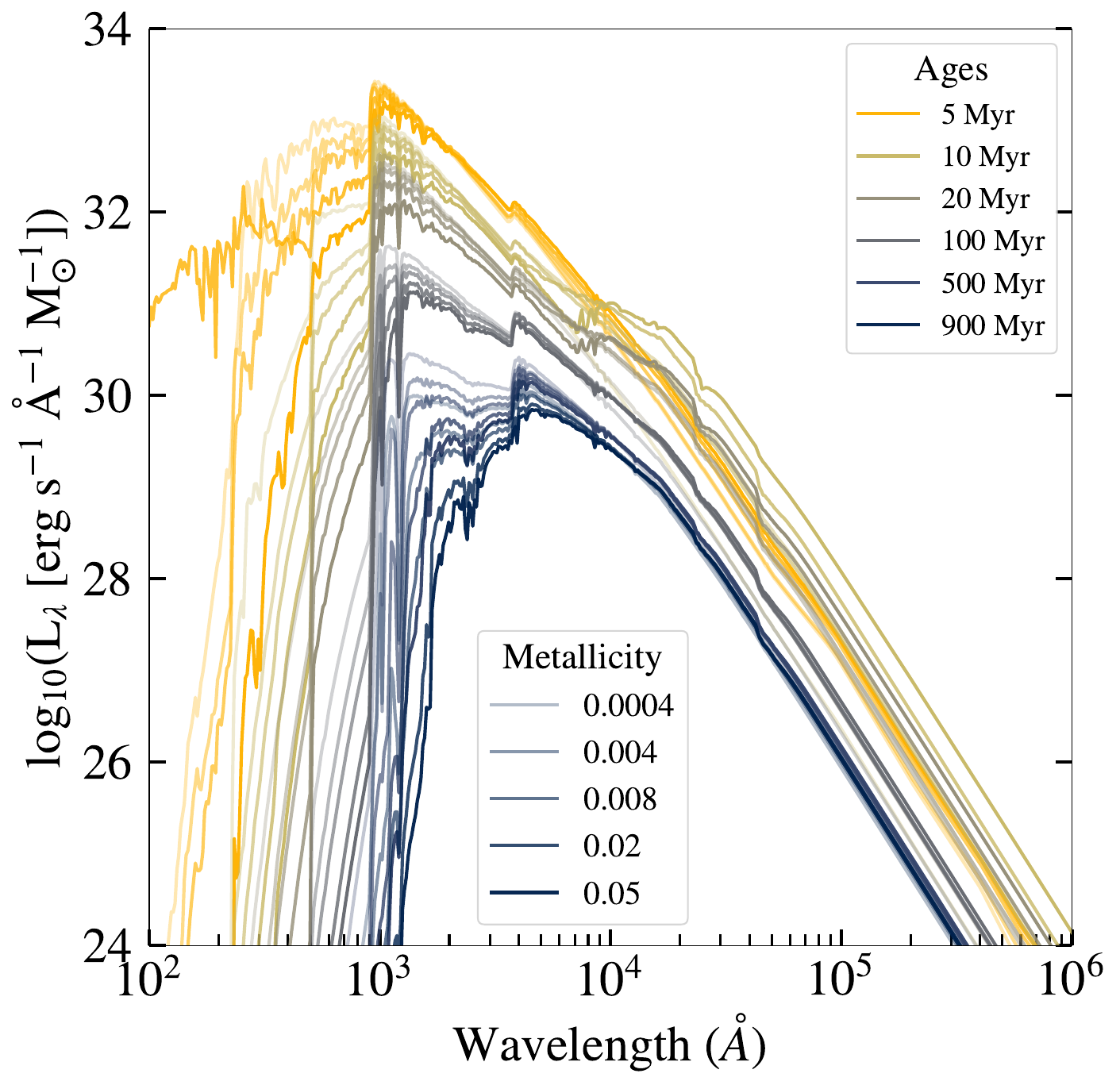}
    \caption{SSP luminosities from the \textsc{Starburst99} code. Populations of different metallicities are shown with different line opacities, whereas the colors represent different ages.
    }
    \label{fig:SSPluminosities}
\end{figure}

In order to study the behavior and evolution of stellar populations, it is necessary to turn to simulations. Several codes have been developed for this purpose and their output spectra $L_{\nu}(t)$ then serve as the input to calculate the stellar emissivities in Eq.~\eqref{eq:emissStellar}. In our case, \textsc{Starburst99}~\cite{leitherer_starburst99_1999,leitherer_effects_2014} has been used to obtain the synthetic stellar spectra in our baseline model. We show them in Fig.~\ref{fig:SSPluminosities}.

We have used a Kroupa IMF, with $M_\mathrm{min}=0.1\,\mathrm{M}_\odot$ and $M_\mathrm{max}=120\,\mathrm{M}_\odot$~\cite{kroupa2002}. We follow the evolutionary Padova tracks.

\subsubsection{The star formation rate \texorpdfstring{$\rho_{\star}$}{rho star}}
The Star Formation Rate (SFR) $\rho_{\star}$ is the amount of SSPs that form at any epoch per unit volume. We follow Ref.~\cite{madau_cosmic_2014} and parametrize the $\rho_{\star}$ as
\begin{equation}\label{eq:sfr}
    \rho_{\star}(z) = x_0 \frac{(1 + z)^{x_1}}{1 + [(1+z)/x_2]^{x_3}}.
\end{equation}
The parameters $x_i$, $i=1,2,3$ are determined by fitting it to the observations. The same approach has been used by, e.g., Refs.~\cite{madau_cosmic_2014, madau_radiation_2017, finke_modeling_2022}.

\subsubsection{Dust escape fraction}
The dust escape fraction $f_\mathrm{esc, dust}$ is the fraction of photons that escape dust absorption.
As we mentioned before, dust absorption plays an important role in the COB emission.
The authors of Ref.~\cite{kneiske_implications_2002} adopt an average fraction of photons that get absorbed at any point in time, disregarding the metallicity evolution. On the other hand, \citetalias{finke_modeling_2022} implements a model relying on explicit redshift dependency, which we also follow here: 
\begin{widetext}\label{eq:dust_Finke22}
\begin{align}
    f_\mathrm{esc, dust}(\lambda, z) =& \frac{f_\mathrm{esc, dust}(\lambda, z=0)}{f_\mathrm{esc, dust}(\lambda=0.15\,\mu m, z=0)} 10^{-0.4 A(z)},\\
    A(z) =& m_d \frac{(1 + z)^{n_d}}{1 + [(1+z)/p_d]^{q_d}},
\end{align}
where $f_\mathrm{esc, dust}$ is a stepwise function,
\begin{align}
    f_\mathrm{esc, dust}(\lambda, z=0) = \begin{cases}
      f_{\mathrm{esc, } 2} + (f_{\mathrm{esc, } 2} - f_{\mathrm{esc, } 1})\cdot(\log_{10}\lambda - \log_{10}\lambda_2)/(\log_{10}\lambda_2-\log_{10}\lambda_1) &\lambda \leq \lambda_2, \\
      f_{\mathrm{esc, } 3} + (f_{\mathrm{esc, } 3} - f_{\mathrm{esc, } 2})\cdot(\log_{10}\lambda - \log_{10}\lambda_3)/(\log_{10}\lambda_3-\log_{10}\lambda_2) &\lambda_2 < \lambda \leq \lambda_3, \\
      f_{\mathrm{esc, } 4} + (f_{\mathrm{esc, } 4} - f_{\mathrm{esc, } 3})\cdot(\log_{10}\lambda - \log_{10}\lambda_4)/(\log_{10}\lambda_4-\log_{10}\lambda_3) &\lambda_3 < \lambda \leq \lambda_4, \\
      f_{\mathrm{esc, } 5} + (f_{\mathrm{esc, } 5} - f_{\mathrm{esc, } 4})\cdot(\log_{10}\lambda - \log_{10}\lambda_5)/(\log_{10}\lambda_5-\log_{10}\lambda_4) &\lambda_4 < \lambda.
    \end{cases}
\end{align}
\end{widetext}

In \citetalias{finke_modeling_2022}, all $\lambda_i, i=1,...,5$ are fixed and the normalizations $f_{\mathrm{esc, dust, }i}$ and the parameters $m_\mathrm{d}$, $n_\mathrm{d}$, $p_\mathrm{d}$, $q_\mathrm{d}$ are fitted to the data.
Here, we use their best-fit parameters, $\lambda_{i} = (0.15, 0.167, 0.218, 0.422, 2.)\,\mu$m, $f_{\mathrm{esc, dust, }i} = (0.188, 0.218, 0.293, 0.393, 0.857)$, and $m_\mathrm{d}=1.49$, $n_\mathrm{d}=0.64$, $p_\mathrm{d}=3.4$, $q_\mathrm{d}=3.54$.
We compare this characterization with the one given by Ref.~\cite{kneiske_implications_2002},

\begin{align}\label{eq:dust_Kneiske02}
    f_\mathrm{esc, dust} = 10&^{-0.4\,A_{\lambda}},\\
    A_{\lambda} = 0.68 \cdot E(B-V) \cdot& R \cdot (\lambda^{-1}-0.35)
\end{align}
with $R$ = 3.2, where $A_{\lambda}$ is the absorption coefficient, and $E(B-V)$ is the color coefficient, which has to be measured and can depend on redshift~\cite{kneiske_implications_2002}. Following Ref.~\cite{kneiske_implications_2002} we neglect the redshift dependence and set $E(B-V)=0.15$.

\subsubsection{Mean metallicity evolution}
The average metallicity is defined as the mean fraction of metals present in an epoch. We follow the parametrization from Ref.~\cite{2022Tanikawa},
\begin{equation} \label{eq:metallEvol}
    Z (z) = Z_\odot \times 10^{a_0 - a_1 \times z ^ {a_2}}
\end{equation}
where $Z_\odot$ is the solar metallicity, which we assume to be $Z_\odot = 0.02$~\cite{2009Zsun}. 
There is a degeneracy between the solar metallicity and $a_0$, and we choose to fix $Z_\odot$ and fit $a_0$.

\subsection{Fit to the data}
In order to break the degeneracy between the axion and stellar components, we include additional data in the COB model fit to constrain it. First, the stellar emissivity is constructed following Eq.~\eqref{eq:emissStellar}. The $\rho_{\star}$ follows the function in Eq.~\eqref{eq:sfr}, and the luminosities come from synthetic stellar spectra. The metallicity of the SSP follows the parametrization of Eq.~\eqref{eq:metallEvol}. 
As \textsc{Starburst99} SSPs are not available for metallicities $< 4\times10^{-4}$, we assume that the SSPs with lower metallicity share the same spectra as this limiting case.
Using the emissivity, we can then calculate the CB spectrum following Eq.~\eqref{eq:nuInu_general}. This gives the CB spectrum of the stellar contribution.

\begin{figure}
    \centering
    \includegraphics[width=0.99\linewidth]{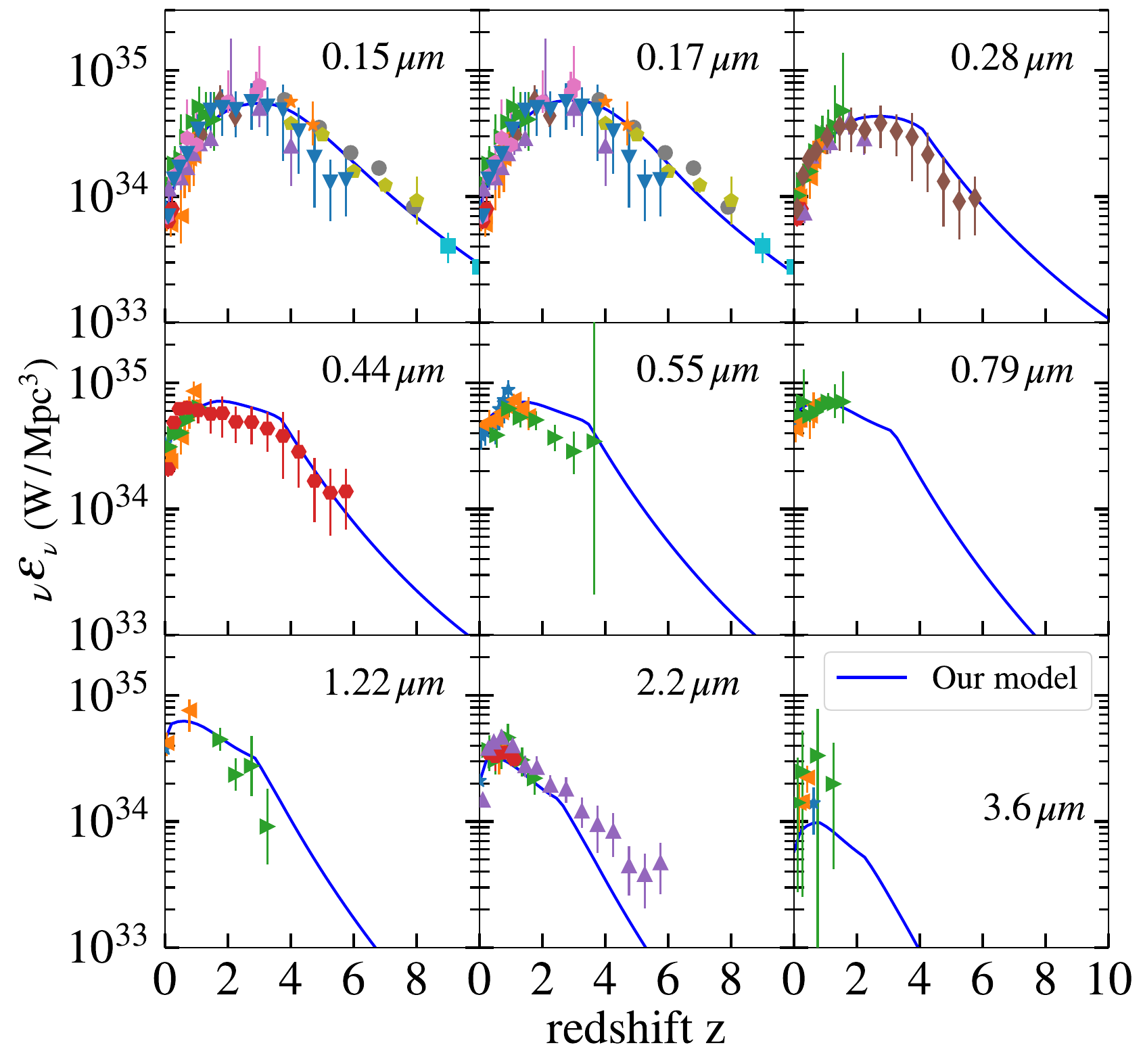}
    \caption{Emissivity data for specific wavelengths as a function of redshift. Observational data is taken from the compilation by \citetalias{finke_modeling_2022}. Solid lines are our best-fit curves.  The $1\,\sigma$ uncertainty band is too thin to be seen.}
    \label{fig:emissivities}
\end{figure}

\begin{figure}
    \centering
    \includegraphics[width=0.99\linewidth]{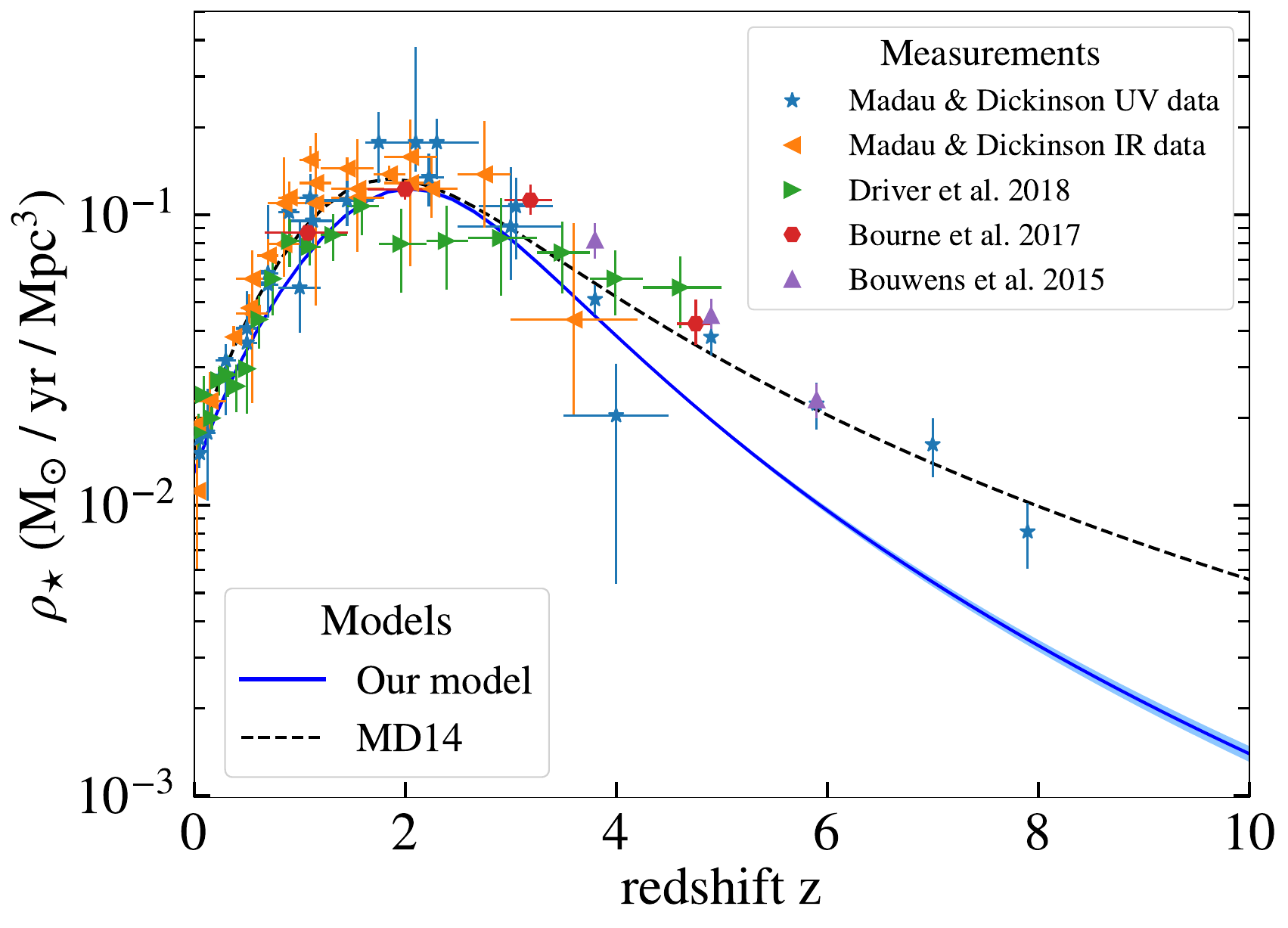}
    \caption{The $\rho_{\star}$ as a function of redshift, with the $1\,\sigma$ uncertainty shown as a light blue band. Observational data compiled by~\cite{the_fermi-lat_collaboration_gamma-ray_2018} are shown with different symbols. The parametrization of Ref.~\cite{madau_cosmic_2014} is shown as a dashed line for comparison. 
    }
    \label{fig:sfr}
\end{figure}

\begin{figure}
    \centering
    \includegraphics[width=0.99\linewidth]{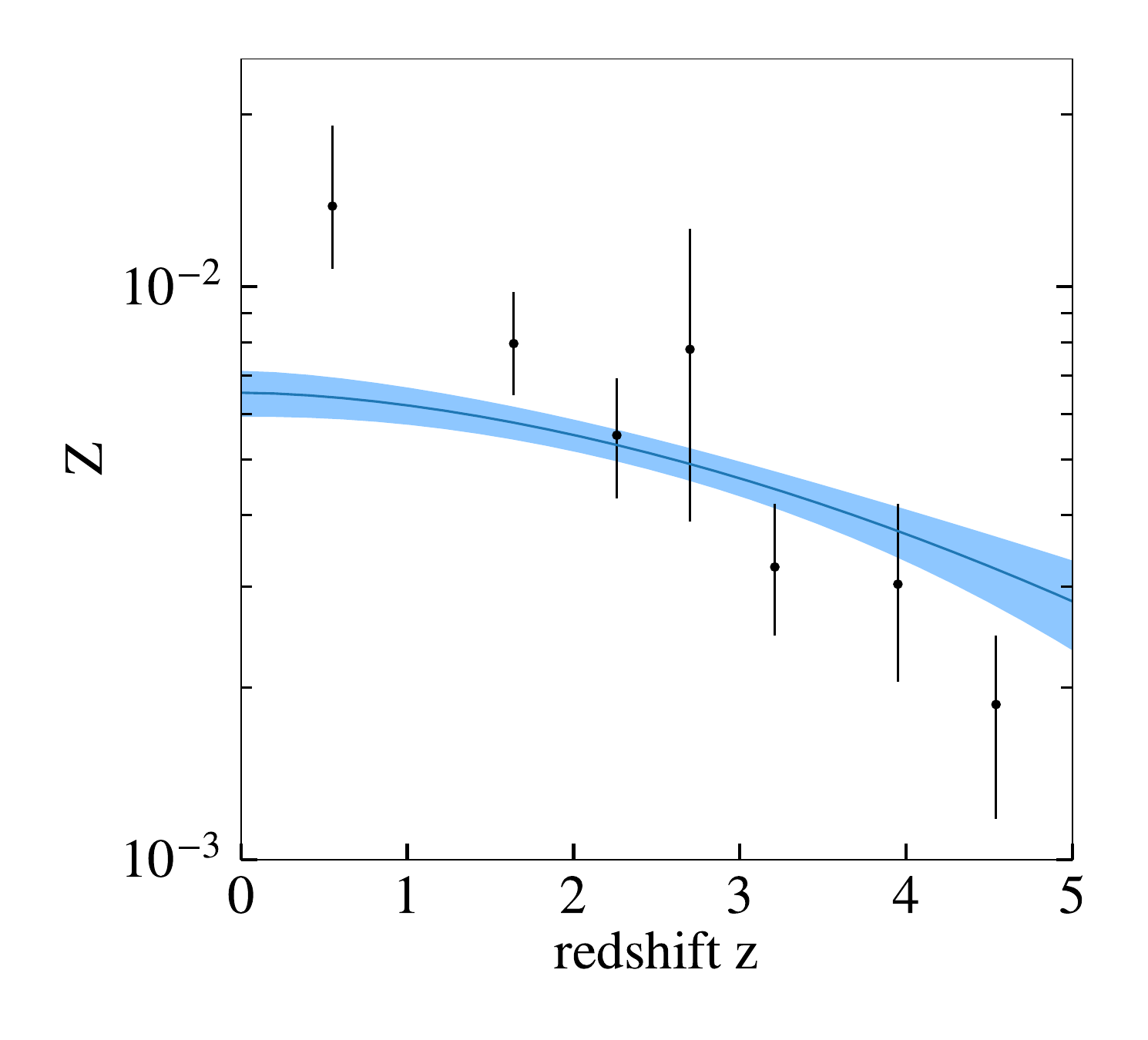}
    \caption{The mean metallicity \textit{Z} as a function of redshift. Observational data taken from Ref.~\cite{2020Zmeasurs}. Best-fit line and $1\,\sigma$ uncertainty are shown as a blue line and blue shaded region.
    }
    \label{fig:metallEvol}
\end{figure}

We use the \textsc{python} package \textsc{iminuit}~\cite{iminuit} in order to fit our model to the observational data, using the least squares method. The methodology is similar to the one used in \citetalias{finke_modeling_2022}. We free the $\rho_{\star}$ and $Z$ parameters and fit our model to the compilations of $\rho_{\star}$ data, the emissivity data, the $Z$ data and the COB galaxy counts that we described in Sec.~\ref{section:measurements}. Emissivities and COB galaxy counts have been calculated from direct galaxy measurements, so they can be used together to constrain the stellar population. This method gives information about the history of stellar formation.
This leads to minimizing the following $\chi^2$

\begin{equation} \label{eq:likelihoodLowerLimits}
\begin{split}
    \chi^2_\mathrm{COB} &= \sum_i \left(\frac{\nu I_{\nu_\mathrm{model}} (\lambda_i) - \nu I_{\nu, i} }{\sigma_i}\right)^2 \\
    &+ \sum_i \left(\frac{\varepsilon_\mathrm{model} (\lambda_i, z) - \varepsilon_i}{\sigma_i}\right)^2 \\
    &+ \sum_i \left(\frac{\rho_{\star, \mathrm{model}} (z_i) - \rho_{\star, i} }{\sigma_i}\right)^2 \\
    &+ \sum_i \left(\frac{Z_{\mathrm{model}} (z_i) - Z_{i} }{\sigma_i}\right)^2.
\end{split}
\end{equation}
The fits of our COB model to observational data are plotted in Fig.~\ref{fig:emissivities} for the emissivities, Fig.~\ref{fig:sfr} for the $\rho_{\star}$, Fig.~\ref{fig:metallEvol} for \textit{Z} and Fig.~\ref{fig:measursOverview} and top panel of Fig.~\ref{fig:cobAndParams} for the COB intensity. Following Eq.~\eqref{eq:likelihoodLowerLimits}, we perform a combined fit of all these data with eight free parameters (the $\rho_{\star}$ and metallicity evolution parameters), which leaves 440 degrees of freedom. We impose no limits to the free parameters.

\subsubsection{Results from the fit}
The reduced $\chi^2$ value for our model is $\chi^2_\mathrm{COB}$ = 2.7. We show the best-fit parameters for our model in Table~\ref{tab:bestfits}. The fit is largely driven by the emissivity data, while the COB and \textit{Z} data make up the smallest contributions to the $\chi^2$. Therefore, the inclusion of $\rho_{\star}$ and emissivity data completely fixes the COB.

At $\lambda \, \scriptstyle {\gtrsim}$$\, 5 \mu$m the dust reemission becomes the dominant component of the CB~\cite{hill_spectrum_2018}, so our fit takes into account CB data up to this wavelength. Our model fails to reproduce the upward trend after this point due to dust, which is visible in the \citetalias{finke_modeling_2022} model in Fig.~\ref{fig:measursOverview}.

\begin{table}
    \centering
    \begin{tabular}{cc} \midrule\midrule
        Parameter & Value \\ \midrule
        \multicolumn{2}{c}{$\rho_{\star}$} \\ \midrule
        $x_0$ $\left(\mathrm{M}_{\odot} / \mathrm{yr} / \mathrm{Mpc}^{3}\right)$ & $(1.302 \pm 0.032) \times 10^{-2}$ \\
        $x_1$ & $2.42\pm 0.04$ \\
        $x_2$ & $3.317\pm 0.035$ \\
        $x_3$ &  $6.70\pm0.07$ \\ \midrule
        \multicolumn{2}{c}{\textit{Z}} \\ \midrule
        $a_0$ & $-0.49\pm0.04$ \\
        $a_1$ &  $0.022\pm0.015$ \\
        $a_2$ & $1.7\pm0.4$
    \end{tabular}
    \caption{Best-fit parameters for our model.}
    \label{tab:bestfits}
\end{table}

\subsection{Systematic uncertainties}
In order to assess the systematic uncertainties with respect to our model choices, we repeat our fits to the COB data with different assumptions. 
First, we used the code \textsc{Pégase} version 3.0.1~\cite{fioc_pegase3_2019} to calculate stellar luminosities. These spectra are similar to those of \textsc{Starburst99}, displaying differences especially at large wavelengths when the power-law regime settles at \\
$\lambda \, \scriptstyle {\gtrsim}$$\, 3 \times 10^4 \mathrm{\AA}$.

We also tried to use the Ref.~\cite{kneiske_implications_2002} dust absorption, instead of the one in \citetalias{finke_modeling_2022}. This prescription ignores any redshift dependence, which affects the lower wavelength regime the most. There is a large amount of absorption at lower wavelengths, which results in higher emissions through the whole spectrum. This however, is at odds with the emissivity data. 

The results from the fits using these different model assumptions are summarized in Table~\ref{tab:chi2_various}. We compare the results for all possible combinations of the considered dust absorption and SSP spectra. Our baseline model with \textsc{Starburst99} spectra and \citetalias{finke_modeling_2022} dust absorption gives the overall best-fit result. Our model has the highest intensity in the LORRI band, and the biggest difference in mean intensity between models is of 14\%, following Eq.~\eqref{eq:meanFlux}.

\begin{table}
    \centering
    \begin{tabular}{ccc}\midrule\midrule
       Model & $\chi^2/\mathrm{d.o.f.}$ & \\ \midrule
        Our model & $1169.2 / 440 = 2.7$ & \\
        \textsc{Pégase} SSPs & $1571.5 / 440 = 3.6$ & \\
        Kneiske dust absorption \cite{kneiske_implications_2002} & $2190.7 / 440 = 5.0$ & \\
        \textsc{Pégase} SSPs and Kneiske dust absorption & $3818.7 / 440 = 8.7$ & \\\midrule
    \end{tabular}
    \caption{$\chi^2$ values over degrees of freedom for the various model combinations described in the text.}
    \label{tab:chi2_various}
\end{table}

\subsection{Additional contributions} \label{appendix:additional}
In addition to light from SSPs, the COB could also originate from other components. Here, we investigate some of them previously discussed in the literature and explain their contribution to our model.

\subsubsection{Intrahalo light} \label{appendix:IHL}

The intrahalo light (IHL) is defined as the light created by stars that have been expelled from their host galaxy due to dynamical events.
We follow the characterization described in Ref.~\cite{bernal_cosmic_2022}, based on calculating the emissivity:
\begin{equation}
     \varepsilon_{\lambda, \mathrm{IHL}} (\lambda, z) = \int ^{M_\mathrm{max}}_{M_\mathrm{min}} dM \, \frac{dn_\mathrm{h}}{dM} \, L_{\lambda, \mathrm{IHL}}(M, z),
\end{equation}
where $dn_\mathrm{h}/dM$ is the halo mass function and $L_{\lambda, \mathrm{IHL}}(M, z)$ is the IHL light specific luminosity emitted by a halo of mass $M$.
We calculate the $dn_\mathrm{h}/dM$ with the \textsc{hmf} code~\cite{202hmfcalculator}, a specific tool for calculating halo mass functions.
We use Ref.~\cite{bernal_cosmic_2022} to characterize $L_{\lambda, \mathrm{IHL}}(M, z)$, taking their definition and reported mean values for the fraction of the total halo luminosity coming from IHL and redshift power-law dependence.

\begin{figure}[h]
    \centering
    \includegraphics[width=0.99\linewidth]{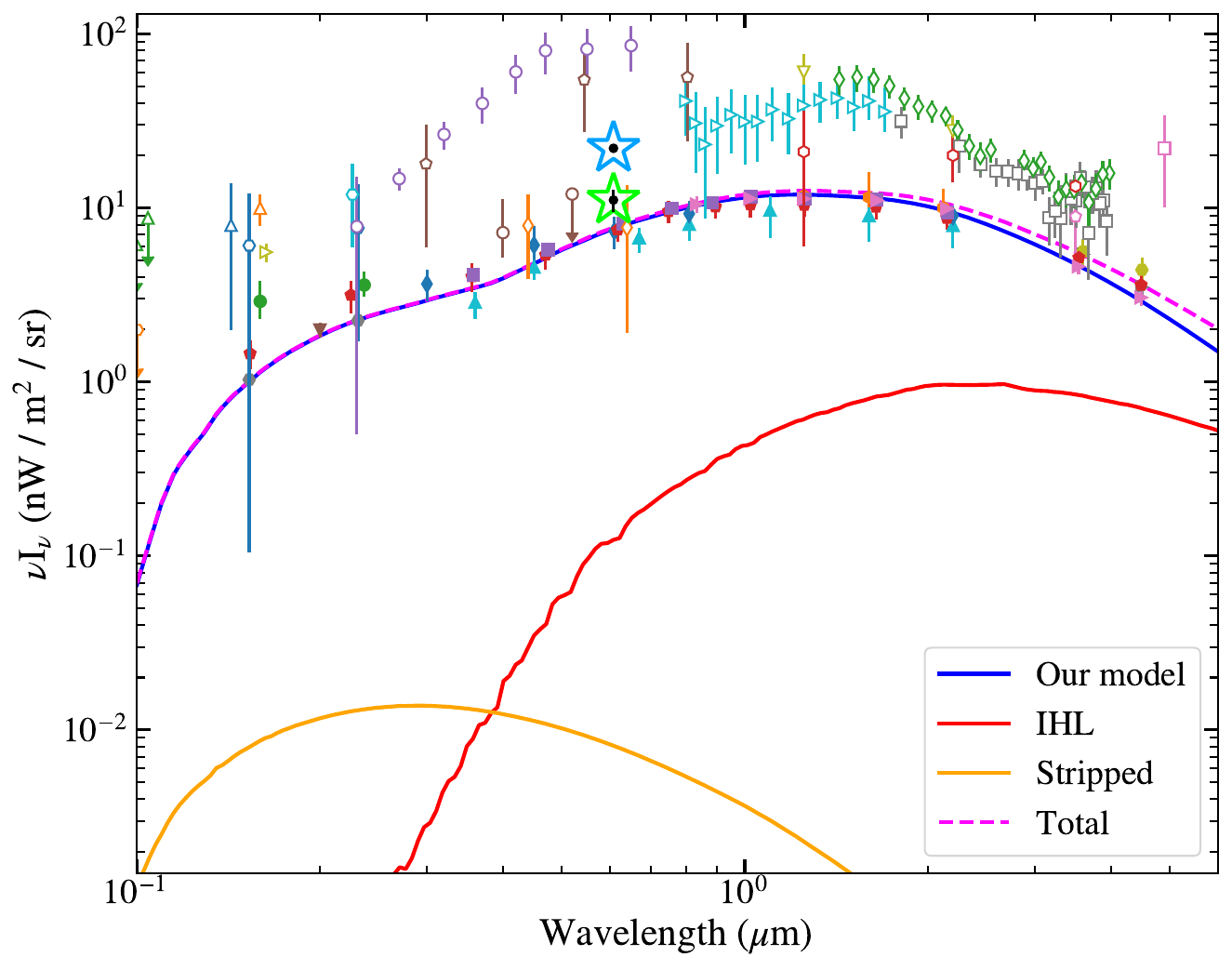}
    \caption{Additional sources of the COB added to our model.}
    \label{fig:cob_IHL}
\end{figure}

Without fitting the COB with the IHL, in Fig.~\ref{fig:cob_IHL} we show the IHL contribution summed to our model. 
The difference to our baseline model is negligible, only becoming relevant at large wavelengths. At $\lambda = 5 \mu$m (the maximum wavelength considered in our analysis), the IHL intensity is about 30\% of the SSP intensity.

In the axion-photon parameter space, the IHL does lower the constraints of lighter axions. However, we have seen in the main text how differences between COB models affect the axion constraints in Fig.~\ref{fig:cobAndParams}. Therefore, the impact of the IHL is negligible.

\subsubsection{Stripped stars}
Stars that strip their envelopes by interacting with a binary companion are defined as stripped. They emit a fraction of their radiation as ionizing photons, with energies larger than the usual stellar evolutionary tracks. 
Therefore, we test whether this extra component could also affect our models. 
We base our calculations on the work of Ref.~\cite{gotberg_y_impact_2019}.
They calculate stellar spectra for these stripped stars, which we include in the COB calculation.
The COB component of this extra contribution is at best 1\% of the COB SSP intensity, as shown in Fig.~\ref{fig:cob_IHL}.

\section{Power-law dependence of axion constraints at higher masses} \label{appendix:powerLaws}
Looking at Fig.~\ref{fig:paramsGITHUB}, we see that there is an almost power-law-like dependence of the constraints on $g_{a \gamma}$ at high masses in each CB region. In this appendix, we derive a relation that approximately describes this behavior. In Eq.~\eqref{eq:axiondecayCOSMIC} we see that
\begin{equation} \label{eq:nuInupropto}
    \nu I_{\nu} (\lambda, z) \propto \frac{(m_a c^2)^2 g_{a\gamma}^2}{\lambda H(z_{\ast})}.
\end{equation}
For high masses corresponding to a decay wavelength shorter than the shortest wavelength where data are available, $\lambda_\mathrm{min}^{\mathrm{C}i\mathrm{B}}\left(i=\mathrm{O,U,X}\right)$, the limits will be dominated by the tail of the axion decay line (due to cosmic expansion) and its contribution to $\nu I_{\nu} \left(\lambda_\mathrm{min}^{\mathrm{C}i\mathrm{B}}\right)$. Considering only this data point, we can solve Eq.~\eqref{eq:nuInupropto} for $g_{a \gamma}$ and use the definition of $z_\star$ from Eq.~\eqref{eq:zstar} to arrive at
\begin{equation}
g_{a\gamma} \propto \frac{\left[H(z_{\ast})\right]^{1/2}}{m_a} =  \frac{\sqrt{H_0}\left[\Omega_{\Lambda} + \Omega_m \left(\frac{m_ac^2}{2hc/\lambda_\mathrm{min}^{\mathrm{C}i\mathrm{B}}}\right)^3\right]^{1/4}}{m_a}.
\end{equation}
Thus, in the limit $m_ac^2 \gg {hc}/{\lambda_\mathrm{min}^{\mathrm{C}i\mathrm{B}}}$,
\begin{equation}
    g_{a\gamma} \propto m_a^{-1/4}.
\end{equation} 
Fitting the slopes of our constraints numerically we find very good agreement with this approximation:
\begin{equation*}
\begin{split}
    m_\mathrm{COB} &= -0.2452 \pm 0.0003 \\
    m_\mathrm{CUB} &= -0.2339 \pm 0.0008 \\
    m_\mathrm{CXB} &= -0.2533 \pm 0.0004,
\end{split}
\end{equation*}
with $g_{a \gamma} \propto  \left(m_ac^2\right)^{m_{\mathrm{C}i\mathrm{B}}}$.

\end{document}